\documentclass[12pt]{article}
\newcommand{\re}{\mathrm{e}}
\newcommand{\ri}{\mathrm{i}}

\usepackage{amsmath,eucal,amssymb}
\usepackage{mathrsfs,graphicx}

\newtheorem{example}{Example}
\newtheorem{remark}{Remark}

\numberwithin{equation}{section}

\def\im{{\mbox{Im}}}
\def\Re{{\mbox{Re}\;}}
\def\ad{\mathrm{ad\,}}
\def\openone{\leavevmode\hbox{\small1\kern-3.3pt\normalsize1}}

\def\bPhi{\mathbf{\Phi}}
\def\bM{\mathbf{M}}
\def\bm{\mathbf{m}}
\def\bbbc{{\Bbb C}}
\def\bbbr{{\Bbb R}}
\def\bbbz{{\Bbb Z}}

\def\diag{\mbox{diag}\,}
\def\tr{\mbox{tr}\,}

\newtheorem{proposition}{Proposition}

\begin{document}

\begin{center}
{\Large \bf On Soliton Interactions for a Hierarchy of \\[2pt]  Generalized Heisenberg
Ferromagnetic Models\\[5pt] on $SU(3)/S(U(1)\times U(2))$ Symmetric Space}

\bigskip

{\bf V. S. Gerdjikov$^1$, G. G. Grahovski$^{1,2}$, A. V. Mikhailov$^3$, \\
T. I. Valchev$^1$}

\medskip

{\it $^1$Institute of Nuclear Research and Nuclear Energy,  Bulgarian Academy of Sciences, \\ 72 Tsarigradsko chausee, Sofia 1784, Bulgaria } \\
{\it $^{2}$School of Mathematical Sciences, Dublin Institute of Technology, \\
Kevin Street, Dublin 8, Ireland} \\
{\it $^3$Applied Math. Department, University of Leeds,\\
Woodhouse Lane, Leeds, LS2 9JT, UK}

\medskip

{\small E-mails: gerjikov@inrne.bas.bg, grah@inrne.bas.bg}\\ {\small a.v.mikhailov@leeds.ac.uk, valtchev@inrne.bas.bg}

\end{center}

\begin{abstract}
\noindent
We consider an integrable hierarchy of nonlinear evolution equations (NLEE) related to linear
bundle Lax operator $L$. The Lax representation is $\bbbz_2\times\bbbz_2$ reduced
and is naturally associated with the symmetric space $SU(3)/S(U(1)\times U(2))$. The simplest
nontrivial equation in the hierarchy is a generalization of Heisenberg ferromagnetic model. We
construct  the $N$-soliton solutions for an arbitrary member of the hierarchy by using the
Zakharov-Shabat dressing method with an appropriately chosen dressing factor.
Two types of soliton solutions: quadruplet  and doublet solitons are found.
The one-soliton solutions of NLEEs with even and odd dispersion laws have
different properties. In particular, the one-soliton solutions for NLEEs
with even dispersion laws are {\em not} traveling waves; their velocities and their amplitudes are time dependent.
Calculating the asymptotics of the $N$-soliton solutions for $t\to\pm\infty$ we analyze
the interactions of quadruplet solitons.
\end{abstract}


\section{Introduction}

The main object of present paper is the following coupled system of equations
\begin{equation}
\begin{split}
\ri u_t &+ u_{xx}+(uu^*_x+vv^*_x)u_x+(uu^*_x+vv^*_x)_x u=0\\
\ri v_t &+ v_{xx}+(uu^*_x+vv^*_x)v_x+(uu^*_x+vv^*_x)_x v=0
\end{split},
\label{2hfe}\end{equation}
where the smooth functions $u:\bbbr^2\to\bbbc$ and $v:\bbbr^2\to\bbbc$
satisfy the algebraic constraint $|u|^2+|v|^2=1$. The system (\ref{2hfe})
is a natural candidate to be a multicomponent generalisation of the classical
Heisenberg ferromagnetic equation. It is well known \cite{blue-bible} that
the Heisenberg ferromagnetic model is integrable in the sense of inverse
scattering method (ISM). It has a Lax pair related to the algebra $\mathfrak{su}(2)$.
Since the time the complete integrability of HF equations was discovered, many
attempts for its generalization have been made \cite{golsok,golsok1,golsok2}. A well
known method \cite{mik_toda,miktetr,mik,mik_ll,GGK*01,Holy*01,lomsan,miklom1,miklom} to
obtain new integrable nonlinear evolution equations (NLEE) is based on imposing
certain algebraic reductions on generic Lax operators. Lax pairs associated to
hermitian symmetric spaces represent a special interest in modern theory of
integrable systems is study of NLEEs \cite{AthFor,For,ForKu*83, JG*15} since the NLEEs
they produce look relatively simple.

The system (\ref{2hfe}) is also integrable in the sense of ISM. Its Lax operators
are associated with the symmetric space $SU(3)/S(U(1)\times U(2))$
with a $\bbbz_2\times\bbbz_2$ reduction imposed on them \cite{ours,ours2,ours3}.

The purpose of the present paper is to derive the soliton solutions
for the integrable hierarchy of equations related to (\ref{2hfe}) and analyse
the interactions between them. In this sense this paper is a natural continuation
of our previous papers \cite{ours,ours2,ours3}.

In Section 2 we start with some basic facts to be used further in paper. Firstly we
describe the hierarchy of nonlinear equations related to (\ref{2hfe}) in terms of
recursion operators. Then we outline the spectral properties of the relevant Lax operator
and formulate direct scattering problem. The spectrum of scattering operator $L$ consists
of a continuous and a discrete parts. As a result of the $\bbbz_2$ reductions $L$ possesses
two configurations of discrete eigenvalues: generic ones, coming in quadruplets $\pm\lambda_k$, $\pm\lambda_k^*$
and purely imaginary ones coming as doublets $\pm \ri\kappa_j$.

In Section 3 we derive the 1-soliton solutions for the NLEEs
of the hierarchy. For this to be done we apply the Zakharov-Shabat dressing method \cite{brown-bible,zakharovshabat,zm1,zm2}
with a rational dressing factor with 2 simple poles. Due to the action of reductions we have two types of 1-soliton solutions:
quadruplet solitons to correspond to 4 eigenvalues and doublet ones to correspond to 2 eigenvalues respectively.
We present explicit expressions for these two types of one-soliton solutions. In order to construct general multisoliton
solutions we discuss two different purely algebraic constructions: by using a
multiple pole dressing factor and by applying "one-soliton" dressing factors several times consecutively.
It turns out that the properties of the 1-soliton solutions to NLEEs with even
and odd dispersion laws differ drastically. For example, the 1-soliton solutions for NLEEs with even dispersion laws are {\em not} traveling waves. Even the doublet soliton of eq. (\ref{2hfe}) exhibits two maxima (resp. minima)
for $|u_1|$ (resp. for $|v_1|$) which first come closer to each other and then move away, one to $\infty$ and
the other to $-\infty$ as time goes to $t\to\infty$. Their velocity, as well as their amplitudes are time dependent.
These properties are similar to the ones of the boomerons and trappons discovered by Calogero and Degasperis
\cite{CaDe*76_1,CaDe*76_2,CaDe1,CaDe2}.
At the same time the soliton solutions to the NLEEs with odd dispersion laws (e.g. the solutions of eq. (\ref{nee2}))
behave as standard solitons, i.e. they are traveling waves.

Section 4 is dedicated to interactions of quadruplet soliton solutions for the NLEE with odd dispersion laws.
In order to do this we use  the classical method of Zakharov
and Shabat, see the monographs \cite{brown-bible,blue-bible} for a detailed exposition.
Namely, we calculate the limits of the $N$-soliton solutions for $t\to\pm\infty$ assuming that all solitons
move with different velocities. In this way we establish that the solitons preserve their velocities and
amplitudes; the only effect of their interaction consists in shifts  of the relative mass center and the phase
of solitons. We provide explicit expressions for these shifts in terms of the poles
$\mu_k$ of the dressing factors.

In Section 5 we briefly discuss the conservation laws of the NLEE and finish with some conclusions.

\section{Preliminaries}

In this section we shall expose in brief some basic facts on Lax operators and direct
scattering problem for the integrable hierarchy of the equation (\ref{2hfe}). In doing
this we shall use a gauge covariant formulation \cite{PhD3,GVY*08,CMP103,GeYa*85}.

\subsection{Polynomial Lax Pair Related to $SU(3)/S(U(1)\times U(2))$}\label{ssec:2.1}

The NLEEs under consideration in this paper represent a zero curvature condition $[L,A] = 0$
for Lax operators $L$ and $A$ in the form:
\begin{eqnarray}
L(\lambda)&=& \ri \partial_x +  \lambda L_1(x,t)\label{lax1} \\
A(\lambda)&=& \ri\partial_t   + \sum^N_{k=1}\lambda^k A_k(x,t),\label{lax2}
\end{eqnarray}
where $\lambda\in\bbbc$ is the so-called spectral parameter and the functions $L_1$
and $A_{k}$, $k=1,\ldots,N$ take values in $\mathfrak{sl}(3,\bbbc)$. The Lax operators
are subject to the following $\bbbz_2$ reductions:
\begin{eqnarray}
L^{\dag}(\lambda^*)&=& -\breve{L}(\lambda)  , \qquad   A^{\dag}(\lambda^*) =
-\breve{A}(\lambda)\label{red1}\\
\mathbf{C} L(-\lambda) \mathbf{C} &=& L(\lambda),\qquad
\mathbf{C} A(-\lambda)\mathbf{C} = A(\lambda),\label{red2}
\end{eqnarray}
where $ {\bf C}=\diag(1,-1,-1)$ and the operation $\breve{}$ is defined as follows
\[\breve{L}(\lambda)\psi(x,t,\lambda) \equiv \ri \partial_x\psi(x,t,\lambda)
 - \lambda \psi(x,t,\lambda) L_1(x,t,\lambda). \]
Due to reduction (\ref{red1}) the matrix coefficients of the Lax pair are hermitian matrices.
On the other hand reduction (\ref{red2}) represents an action of Cartan's involutive automorphism
which defines the symmetric space $SU(3)/S(U(1)\times U(2))$, see \cite{Hel,loos}.
It induces a $\bbbz_2$-grading in the Lie algebra $\mathfrak{sl}(3,\bbbc)$
\begin{equation}\label{grading}
\mathfrak{sl}(3) = \mathfrak{sl}^{0}(3)\oplus\mathfrak{sl}^{1}(3),
\qquad\mathfrak{sl}^{\sigma}(3) = \{ X\in \mathfrak{sl}(3) \, | \,
 \mathbf{C} X \mathbf{C} = (-1)^{\sigma} X\}.
\end{equation}
It is evident that $L_1, A_{k}\in\mathfrak{sl}^1(3)$ for $k$ being an odd integer
and $A_{k} \in \mathfrak{sl}^0(3)$ otherwise. This means that $A_k$ for even $k$ are
block-diagonal matrices of the form
\[A_k = \left(\begin{array}{ccc}
\ast & 0 & 0 \\ 0 & \ast & \ast  \\
0 & \ast & \ast
\end{array}\right) \]
while $L_1$ and $A_k$ for odd $k$ have the complementary block structure. In particular,
$L_1$ is written as:
\begin{equation}
L_1 = \left(\begin{array}{ccc}
0 & u & v \\ u^* & 0 & 0 \\
v^* & 0 & 0
\end{array}\right)
\end{equation}
The potential $L_1$ is required to obey the following conditions:
\begin{enumerate}
\item  The eigenvalues of $L_1$ are $0, \pm 1$, i.e. the potential satisfies the characteristic
equation $L^3_1=L_1$.
\item The function $L_1(x,t) - L_{\pm}$ where
\begin{equation}
\lim_{x\to \pm\infty} L_1(x,t) = L_{\pm} = \left(\begin{array}{ccc} 0 & 0 & \re^{\ri \phi_\pm}\\
0 & 0 & 0 \\ \re^{-\ri \phi_\pm} & 0 & 0\end{array}\right),\qquad \phi_{\pm}\in\bbbr.
\label{const_bc}\end{equation}
is a Schwartz type function, i.e. it is infinitely smooth and tends to $0$ faster
than any polynomial when $|x|\to\infty$.
\end{enumerate}

The grading (\ref{grading}) means that any function $X$ with values in $\mathfrak{sl}(3)$
can be split as follows:
\begin{equation}
X = X^{0} + X^{1}, \qquad X^{0,1}\in\mathfrak{sl}^{0,1}(3).
\label{X_split}\end{equation}
Let us define the Killing form for $\mathfrak{sl}(3)$ as follows:
\[\langle X, Y\rangle = \tr (XY),\qquad X,Y\in\mathfrak{sl}(3).\]
Then each component $X^{0,1}$ splits into a term commuting with
$L_1$ and its orthogonal complement with respect to the Killing form
\begin{eqnarray}
X^{0} &=&   X^{0,\bot} + \kappa_0 L_2,\qquad L_2 = L^2_1 - \frac{2}{3}\openone,
\qquad \langle X^{0,\bot}, L_2\rangle = 0\label{X0_split}\\
X^{1} &=&   X^{1,\bot} + \kappa_1 L_1,\qquad \langle X^{1,\bot}, L_1\rangle = 0.
\label{X1_split}\end{eqnarray}
As a simple consequence of condition 1 above $L_1$ and $L_2$ are normalized as follows:
\begin{equation}
\langle L_1, L_1\rangle = 2,\qquad \langle L_2, L_2\rangle = \frac{2}{3}.
\end{equation}
Therefore the coefficients $\kappa_0$ and $\kappa_1$ are given by the following
equalities
\begin{equation}
\kappa_0 = \frac{3}{2}\langle X^{0}, L_2\rangle,
\qquad \kappa_1 = \frac{1}{2}\langle X^{1}, L_1\rangle.
\label{coef01}\end{equation}

The zero curvature condition $[L,A] = 0$ for the pair (\ref{lax1}), (\ref{lax2}) leads to
certain recurrence relations for the matrix coefficients of $L$ and $A$, see \cite{ours3}.
Resolving them allows one to express $A_k$ in terms of $L_1$ and its $x$-derivatives of order
up to $N-k$. Since the maximal order term in the operator $A$ must commute with $L_1$ there exists two
options:
\[\begin{aligned}
\mbox{a)}\   &   A_N = c_{2p} L_2,\quad &\mbox{if}& \quad N =2p \\
\mbox{b)} \  &   A_N = c_{2p+1} L_1,\quad &\mbox{if}& \quad N=2p+1,
\end{aligned}\]
where $c_{2p}$ and $c_{2p+1}$ are constants. Then a more detailed analysis \cite{ours3} shows that the NLEEs look as follows:
\begin{equation}
\begin{aligned}
&\mbox{a)} &  \ad^{-1}_{L_1}L_{1,t} - \sum_{q=1}^{p} c_{2q} (\Lambda_1\Lambda_2)^{q-1} \Lambda_1 \ad_{L_1}^{-1} L_{2,x}
- \sum_{q=0}^{p-1} c_{2q +1}(\Lambda_1\Lambda_2)^{q} \ad_{L_1}^{-1} L_{1,x}  &=0, \\
&\mbox{b)} &  \ad^{-1}_{L_1}L_{1,t}  - \sum_{q=1}^{p} c_{2q} (\Lambda_1\Lambda_2)^{q-1} \Lambda_1 \ad_{L_1}^{-1}L_{2,x}
-  \sum_{q=0}^{p} c_{2q+1} (\Lambda_1\Lambda_2)^{q}\ad_{L_1}^{-1} L_{1,x}   &=0.
\end{aligned}\label{nlee}
\end{equation}
The integro-differential operators $\Lambda_1$ and $\Lambda_2$ appeared above are
given by
\begin{equation}
\begin{split}
\Lambda_1 = -\ri \ad^{-1}_{L_1} \left(\pi\partial_x(\cdot)
-\frac{1}{2}L_{1,x}\partial^{-1}_x\langle\partial_x(\cdot),
L_1\rangle\right)\\
\Lambda_2 = -\ri \ad^{-1}_{L_1}\left(\pi\partial_x(\cdot\,) -\frac{3}{2}L_{2,x}\partial^{-1}_x
\langle \partial_x (\cdot\,), L_2 \rangle\right),
\end{split}\end{equation}
where projection $\pi:=\ad^{-1}_{L_1}\ad_{L_1}$ cuts all $L_1$-commuting parts off.
The operator
\[ \Lambda X:= \left\{\begin{array}{cc}
\Lambda_1\Lambda_2X ,& X\in\mathfrak{sl}^{0}(3)\\
\Lambda_2\Lambda_1X ,& X\in\mathfrak{sl}^{1}(3)\end{array}\right.\]
is called recursion operator. It can be viewed as an adjoint representation
of the operator $L$. Its existence manifests the hierarchies associated with
NLEE (nonlinear equations, integrals of motion, simplectic forms etc) and thus
plays a very important role in theory of solitons.

\begin{example}
Consider the simplest case when $N=2$. Then the matrix coefficients of the
second Lax operator $A$ read:
\begin{eqnarray}
A_2 &=& - \left(\begin{array}{ccc}
1/3 & 0 & 0 \\ 0 & |u|^2-2/3 & u^*v \\
0 & v^*u & |v|^2-2/3
\end{array}\right),\quad
A_1 = \left(\begin{array}{ccc}
0 & a & b \\ a^* & 0 & 0 \\
b^* & 0 & 0
\end{array}\right)\label{matr_coef1}\\
a&=& \ri u_{x}+\ri (uu^*_x+vv^*_x)u,\quad
b = \ri v_{x}+\ri (uu^*_x+vv^*_x)v
\label{matr_coef2}\end{eqnarray}
This $L$-$A$ pair produces the 2-component system
\begin{equation}
\begin{split}
\ri u_t &+ u_{xx}+(uu^*_x+vv^*_x)u_x+(uu^*_x+vv^*_x)_x u=0\\
\ri v_t &+ v_{xx}+(uu^*_x+vv^*_x)v_x+(uu^*_x+vv^*_x)_x v=0
\end{split}.
\label{nee}\end{equation}
we started our paper with (see (\ref{2hfe})).$\Box$
\end{example}

For completeness here we present another member of the hierarchy (\ref{nlee}).
It is the simplest NLEE corresponding to an odd dispersion law.
\begin{example}
Consider the case when $\mathbf{f}(\lambda) = -8\lambda^3 J$, i.e.
$c_3=-8$, $c_2=c_1=0$. Then the corresponding 2-component system
obtains the form:
\begin{equation}
\begin{split}
u_t &= 8u_{xxx} + 12(uu^*_x+vv^*_x)u_{xx} + r(u,v) u_x
+ s(u,v)u\\
v_t &= 8v_{xxx} + 12(uu^*_x+vv^*_x)v_{xx} + r(u,v) v_x
+ s(u,v)v
\end{split},
\label{nee2}\end{equation}
where
\[\begin{split}
r(u,v)&= 3\left[4(|u_x|^2 + |v_x|^2) + 5(uu^*_x+vv^*_x)^2
+ 6(uu^*_x+vv^*_x)_x \right]\\
s(u,v)& = 3\left[2(uu^*_x+vv^*_x)_{xx} + 4(|u_x|^2 + |v_x|^2)_x
+ 5 (uu^*_x+vv^*_x)^2_x\right]. \quad\Box
\end{split}\]
\end{example}

Sometimes it is more convenient to deal with Lax operators
written in canonical gauge. In this gauge the operator
(\ref{lax1}) looks as follows:
\begin{equation}
\tilde{L}(\lambda) =g^{-1}Lg = \ri \partial_x + U_0(x,t) + \lambda J,
\qquad J=\diag(1,0,-1), \label{lax_1_g}
\end{equation}
where
\begin{equation}
g = \frac{\sqrt{2}}{2}\left(\begin{array}{ccc}
1 & 0 & -1 \\ u^* & \sqrt{2} v & u^*\\
v^* & -\sqrt{2} u  & v^*
\end{array}\right).
\label{g_trans}\end{equation}
The second Lax operator (\ref{lax2}) is given by
\begin{equation}
\begin{split}
\mbox{a)} &\quad \tilde{A}(\lambda) = \ri \partial_t + \sum^{N-1}_{k=0}\lambda^k\tilde{A}_k(x,t) + c_N\lambda^NI,
\qquad N=2p\\
\mbox{b)} &\quad \tilde{A}(\lambda) = \ri \partial_t + \sum^{N-1}_{k=0}\lambda^k\tilde{A}_k(x,t) + c_{N}\lambda^NJ,
\quad \qquad N=2p+1,
\end{split}\label{lax_2_g}\end{equation}
where $I=g^{-1}L_2\,g=\diag(1/3,-2/3,1/3)$.

\subsection{Direct Scattering Problem}\label{ssec:2.2}

In order to formulate a direct scattering problem for $L$, one needs to
introduce auxiliary spectral linear system
\begin{equation}
L(\lambda)\psi(x,t,\lambda) = \ri \partial_x \psi(x,t,\lambda)
+ \lambda L_1(x,t) \psi(x,t,\lambda) = 0.
\label{lax_pol}\end{equation}
Here $\psi$ denotes a fundamental set of solutions or a fundamental solution for short.
Since the operators (\ref{lax1}) and (\ref{lax2}) commute $\psi$ also satisfies
\begin{equation}
A(\lambda)\psi(x,t,\lambda) = \left(\ri \partial_t
+ \sum^N_{k=1} \lambda^k A_k(x,t)\right)\psi(x,t,\lambda)
= \psi(x,t,\lambda) \mathbf{f}(\lambda)
\label{auxsys_2}\end{equation}
as well. The matrix-valued function
\begin{equation}\label{f_lambda}
\mathbf{f}(\lambda)=\lim_{x\to\pm \infty} g_{\pm}^{-1}
\sum^N_{k=1}\lambda^k A_k(x,t)g_{\pm}
\end{equation}
is called dispersion law of the nonlinear equation (\ref{nlee}). The
unitary matrix
\[g_{\pm} = \lim_{x\to\pm\infty}g(x,t) =
\frac{1}{\sqrt{2}}\left(\begin{array}{ccc}
1 & 0 & -1 \\ 0 & \sqrt{2}\,\re^{\ri \phi_\pm} & 0 \\
\re^{-\ri \phi_\pm} & 0 & \re^{-\ri \phi_\pm}
\end{array}\right)\]
involved in the definition of the dispersion law diagonalizes the asymptotics $L_{1,\pm}=\lim_{x\to\pm\infty}L_1(x,t)$.
It can be proven that the dispersion law of (\ref{nlee}) reads
\begin{equation}
\begin{aligned}
&\mbox{a)} &\quad \mathbf{f}(\lambda) & = \sum_{q=0}^{p-1} c_{2q+1} \lambda^{2q+1}J +
\sum_{q=1}^{p} c_{2q} \lambda^{2q}I, \\
&\mbox{b)} &\quad \mathbf{f}(\lambda) & = \sum_{q=0}^{p} c_{2q+1} \lambda^{2q+1}J
 + \sum_{q=1}^{p} c_{2q} \lambda^{2q}I.
\end{aligned}
\label{eq:flam}\end{equation}
The dispersion law of the 2-component system (\ref{nee}) is $ -\lambda^2I$ and
that of ((\ref{nee2}) is $-8\lambda^3 J$. It is evident from (\ref{eq:flam}) that $\mathbf{f}(\lambda)$ obeys the splitting:
\begin{equation}
\mathbf{f}(\lambda) = \mathbf{f}_0(\lambda) I + \mathbf{f}_1(\lambda) J,
\label{f_split}\end{equation}
which is a result of the $\bbbz_2$ grading (\ref{grading}) of the algebra
$\mathfrak{sl}(3)$.

A special type of fundamental solutions are the so-called Jost
solutions $\psi_{\pm}$ which are normalized as follows
\begin{equation}
\lim_{x\to\pm\infty}\psi_{\pm}(x,t,\lambda)
\re^{-\ri \lambda J x}g_{\pm}^{-1} =\openone.
\label{josts}\end{equation}
Due to (\ref{f_lambda}) one can show that the asymptotic behavior of
$\psi_{\pm}$ do not depend on time and thus the definition is correct.
The transition matrix
\begin{equation}
T(t,\lambda) = [\psi_{+}(x,t,\lambda)]^{-1}\psi_{-}(x,t,\lambda).
\label{tmatrix}\end{equation}
is called scattering matrix. It can be easily deduced from relation
(\ref{auxsys_2}) that the scattering matrix evolves with time according
to the linear differential equation
\begin{equation}
\ri \partial_t T+[\mathbf{f}(\lambda),T]=0,
\end{equation}
which is integrated straight away to give
\begin{equation}
T(t,\lambda)=\re^{\ri \mathbf{f}(\lambda)t}T(0,\lambda)\re^{-\ri \mathbf{f}(\lambda)t}.
\end{equation}
From now on the parameter $t$ will be fixed and we shall omit it to simplify
our notation. Due to reasons of simplicity we set $\phi_{+}=\phi_{-} = 0$ as
well.

The action of $\bbbz_2$-reductions (\ref{red1}), (\ref{red2}) imposes
the following restrictions
\begin{equation}
\begin{split}
\left[\psi^\dag _{\pm}(x,\lambda^*)\right]^{-1}
&= \psi_{\pm}(x,\lambda),\qquad
\left[T^\dag (\lambda^*)\right]^{-1} = T(\lambda) \\
\mathbf{C}\psi_{\pm}(x,-\lambda)\mathbf{C} &= \psi_{\pm}(x,\lambda),\qquad
\mathbf{C}T(-\lambda)\mathbf{C} = T(\lambda)
\end{split}\label{jostred}\end{equation}
on the Jost solutions and the scattering matrix.

The continuous spectrum of $L$ fills up the real axis in the complex $\lambda$-plane.
Thus the $\lambda$-plane is divided into two regions denoted by $\bbbc_{+}$ (the upper half
plane) and $\bbbc_{-}$ (the lower half plane). These regions represent domains for fundamental solutions $\chi^{+}(x,\lambda)$ and $\chi^{-}(x,\lambda)$ to be analytic functions in $\bbbc_{+}$ and $\bbbc_{-}$ respectively \cite{ours2}. The fundamental analytic solutions (FAS) can be constructed by using Gauss factors in the decomposition of the scattering matrix:
\begin{equation}\label{eq:T}
T(\lambda)=T^{\mp}(\lambda)D^{\pm}(\lambda)(S^{\pm}(\lambda))^{-1}.
\end{equation}
$S^{+}$ and $T^{+}$ are upper triangular matrices, $S^{-}$ and $T^{-}$
are lower triangular matrices and $D^{\pm}$ are diagonal ones.
Then $\chi^{+}$ and $\chi^{-}$ are expressed as follows
\begin{equation}\label{chi_pm}
\chi^{\pm}(x,\lambda)=\psi_{-}(x,\lambda)S^{\pm}(\lambda)
=\psi_{+}(x,\lambda)T^{\mp}(\lambda)D^{\pm}(\lambda).
\end{equation}
Due to relation (\ref{chi_pm}) the FAS can be interpreted as solutions to a local Riemann-Hilbert problem
\begin{equation}
 \chi^{+}(x,\lambda)=\chi^{-}(x,\lambda)G(x,\lambda),\qquad   G(\lambda) =
(S^-(\lambda))^{-1}S^+(\lambda).
\end{equation}
The established interrelation between the inverse scattering method and
Riemann-Hilbert problem plays an important role in constructing solutions
to NLEEs through dressing method.

It can be shown that the reduction conditions (\ref{jostred})
and equation (\ref{eq:T}) lead to the following demands on the Gauss factors
\begin{equation}\label{eq:TGred}
\begin{aligned}
\left[S^+(\lambda^*)\right]^\dag &= [S^-(\lambda)]^{-1}, &  \qquad  \tilde{\mathbf{C}}S^\pm (-\lambda) \tilde{\mathbf{C}} &= S^\mp (\lambda) \\
\left[T^+(\lambda^*)\right]^\dag &= [T^-(\lambda)]^{-1}, &  \qquad \tilde{\mathbf{C}}T^\pm (-\lambda)\tilde{\mathbf{C}} &= T^\mp (\lambda)\\
\left[D^+(\lambda^*)\right]^\dag &= [D^-(\lambda)]^{-1}, &  \qquad  \tilde{\mathbf{C}}D^\pm(-\lambda)\tilde{\mathbf{C}} &= D^\pm(\lambda),
\end{aligned}
\end{equation}
where
\[\tilde{\mathbf{C}} = \left(\begin{array}{ccc} 0 & 0 & 1 \\ 0 & 1 & 0 \\
1 & 0 & 0  \end{array}\right).\]
Finally, combining all this information we see that the FAS obey the symmetry
conditions
\begin{equation}\label{eq:FAS-red}
\left[\chi^{+}(x,\lambda^*)\right] = [\chi^{-}(x,\lambda)]^{-1} \qquad
\mathbf{C}\chi^{+}(x,-\lambda)\mathbf{C} = \chi^{-}(x,\lambda).
\end{equation}

\section{Dressing Method and Soliton Solutions}

As we mentioned in the previous section the inverse scattering method
is tightly related to Riemann-Hilbert problem. The Riemann-Hilbert
problem possesses two types of solutions: regular ones (without singularities)
and singular ones. Singular solutions can be generated by dressing regular solutions
with a factor which has prescribed singularities. The simplest types of
singularities are first order poles and zeroes. It can be proven that they
correspond to poles of the resolvent of $L$. Hence they are discrete eigenvalues
of the Lax operator (\ref{lax1}). The discrete eigenvalues of $L$ form orbits of
the reduction group $\bbbz_2\times\bbbz_2$. There exist two types of orbits:
generic orbits containing quadruplets of eigenvalues $\{\pm\mu,\pm \mu^*\}$
and degenerate orbits consisting of two imaginary eigenvalues $\pm\ri\kappa$
(doublets).

There is a very deep connection between singular solutions to Riemann-Hilbert
problem and soliton solutions to the corresponding nonlinear problem. In the
present section we are going to analyze the soliton solutions to the
system (\ref{nlee}). For this to be done, we are going to apply the dressing
method proposed in \cite{zakharovshabat} and developed in \cite{zm1,zm2,mik,mik_ll}.
We demonstrate that the NLEE (\ref{nlee}) has two types of $1$-soliton solutions:
doublet soliton to be connected with two imaginary discrete eigenvalues of $L$ and
quadruplet soliton connected to 4 eigenvalues.

\subsection{Rational Dressing}

The dressing method is an indirect method for solving a NLEE possessing a
Lax representation. This means that it allows one to generate a solution to
the NLEE starting from a known one.
Let us assume we know a solution
\[L^{(0)}_{1} = \left(\begin{array}{ccc}
0 & u_0 & v_0 \\ u^*_0 & 0 & 0\\
v^*_0 & 0 & 0\end{array}\right)\]
of (\ref{nlee}) and a fundamental solution $\psi_0(x,t,\lambda)$ of the
auxiliary linear problems
\begin{equation}\begin{split}
L^{(0)}(\lambda)\psi_0 &= \ri\partial_x\psi_0 + \lambda L^{(0)}_1\psi_0 = 0\\ A^{(0)}(\lambda)\psi_0 &= \ri\partial_t\psi_0 +\sum^N_{k=1}\lambda^k A^{(0)}_k
\psi_0 = 0.
\label{bare_lax}
\end{split}\end{equation}
Then one constructs another function $\psi_1(x,t,\lambda)= \Phi(x,t,\lambda)\psi_0(x,t,\lambda)\Phi_-^\dag(\lambda)$, where
$\Phi_-(\lambda)=\lim_{x\to -\infty} \Phi(x,t,\lambda)$. This function is a  common solution to
\begin{equation}
\begin{split}
L^{(1)}(\lambda)\psi_1 &= \ri\partial_x\psi_1 + \lambda L^{(1)}_1\psi_1 = 0\\ A^{(1)}(\lambda)\psi_1 &= \ri\partial_t\psi_1 +\sum^N_{k=1}\lambda^k A^{(1)}_k
\psi_1 = 0,
\label{dressed_lax}\end{split}
\end{equation}
where the potential
\[L^{(1)}_{1} = \left(\begin{array}{ccc}
0 & u_1 & v_1 \\ u^*_1 & 0 & 0\\
v^*_1 & 0 & 0\end{array}\right)\]
is to be found. From  (\ref{bare_lax}) and (\ref{dressed_lax}) it follows that
the dressing factor $\Phi (x,t,\lambda)$ satisfies the following equations:
\begin{eqnarray}\label{eq:dress-eq}
&&\ri \partial_x \Phi + \lambda L_1^{(1)}\Phi
- \lambda\Phi L_1^{(0)} = 0\\
\label{eq:dress-eqA}
&& \ri\partial_t \Phi + \sum^N_{k=1}\lambda^k A_k^{(1)}\Phi
- \Phi \sum^N_{k=1}\lambda^k A_k^{(0)} = 0.
\end{eqnarray}

We also assume that the dressing factor is regular at $|\lambda|\to 0,\infty$.
Then from (\ref{eq:dress-eq}) one can derive the following relation
between $L_1^{(1)}$ and $L_1^{(0)}$:
\begin{equation}\label{eq:dress-tr}
L_1^{(1)}
(x,t)=\Phi(x,t,\infty)L_1^{(0)}(x,t)\Phi^\dag(x,t,\infty).
\end{equation}
This equation will play a central role in our further considerations
since it allows one to generate a new solution to (\ref{nlee})
from the given one $L^{(0)}_1$.

Due to the reduction conditions (\ref{red1}), (\ref{red2}) the dressing
factor obeys the symmetries:
\begin{eqnarray}\label{eq:fas-invol}
&&{\bf C}\Phi (x,t,-\lambda){\bf C}=\Phi(x,t,\lambda)\\
&&\Phi (x,t,\lambda)\Phi^\dag (x,t,\lambda^*)=\openone .
\label{eq:her-invol}\end{eqnarray}

In order to obtain a nontrivial dressing we choose $\Phi(x,t,\lambda)$
as a rational function\footnote{If $\Phi$ is $\lambda$-independent then
it does not depend on $x$ and $t$ either. Thus (\ref{eq:dress-tr}) produces
simply a unitary transformation of $L^{(0)}_1$ which is not essential because
of $U(2)$ gauge symmetry of the model.} of $\lambda$ with a minimal number
of simple poles. At first we shall consider the case when these poles are generic
complex numbers. Hence the dressing factor looks as follows:
\begin{equation}
\Phi(x,t,\lambda)=\openone + \frac{\lambda M(x,t)}{\lambda - \mu} + \frac{\lambda {\bf
C}M(x,t){\bf C}}{\lambda + \mu},
\label{eq:dress-anz1}\end{equation}
where $\Re\mu\neq 0$, $\text{Im}\;\mu\neq 0$. It is evident that the reduction condition (\ref{eq:fas-invol}) is fulfilled. On the other hand (\ref{eq:her-invol}) leads
to the conclusion that
\begin{equation}
\Phi^{-1}(x,t,\lambda)= \openone + \frac{\lambda M^\dag(x,t)}{\lambda - \mu^*}
+\frac{\lambda {\bf C}M^\dag(x,t){\bf C}}{\lambda + \mu^*}.
\label{eq:dress-anz2}\end{equation}

The identity $\Phi(\lambda)\Phi^{-1}(\lambda)=\openone$ must hold for any
$\lambda$. Therefore after equating the residue at $\lambda=\mu^*$ to 0 one
gets the equation:
\begin{equation}\label{eq:dress-m}
\left(\openone+\frac{\mu^* M(x,t)}{\mu^*- \mu} +\frac{\mu^* {\bf
C}M(x,t){\bf C}}{\mu^* + \mu}\right)M^\dag(x,t)=0.
\end{equation}
The rest of algebraic relations can be reduced to (\ref{eq:dress-m}) due to
the symmetry conditions (\ref{red1}), (\ref{red2}).

The residue $M$ ought to be singular since otherwise it should be proportional
to $\openone$ and the dressing becomes trivial. It suffices to consider the
case ${\rm rank} M = 1$. Then $M$ can be decomposed in the following
manner:
\begin{equation}
M=|n\rangle\langle m|,\qquad
|n\rangle=\left(n_1,n_2,n_3\right)^T,\
\langle m|=(m^*_1,m^*_2,m^*_3).
\label{M_decomp}\end{equation}
After substituting this representation into (\ref{eq:dress-m}) one derives a linear
system for the $3$-vector $|n\rangle$:
\begin{eqnarray}\label{eq:n}
|m\rangle - \frac{\mu^* |n\rangle\langle m|m\rangle}{2\ri\kappa}
+ \frac{\mu^* {\bf C} | n\rangle\langle m| {\bf C}| m\rangle}
{2\omega}=0.
\end{eqnarray}
where we have used the notation $\omega = \Re {\mu}$, $\kappa = \text{Im}\;{\mu}$.
The solution of (\ref{eq:n}) reads:
\begin{equation}\label{n}
| n \rangle= \frac{1}{\mu^*}\left(\frac{\langle m|m\rangle}{2\ri\kappa}
- \frac{\langle m|\mathbf{C}|m\rangle}{2\omega}\mathbf{C}\right)^{-1}
| m \rangle.
\end{equation}
The vector $| m \rangle$ is an element of the projective
space ${\Bbb CP}^2$. Indeed, it is evident that a rescaling
$| m \rangle \to h| m \rangle$ with any complex
$h\neq 0$ does not change the matrix $M$.

Taking into account the ansatz (\ref{eq:dress-anz1}) one can
rewrite (\ref{eq:dress-tr}) as:
\begin{equation}\label{L_1}
L_1^{(1)}=(\openone+M+{\bf C}M{\bf C})L_1^{(0)}
(\openone+M+{\bf C}M{\bf C})^\dag.
\end{equation}
Notice that the dressing procedure preserves the matrix structure of $L$ since
the factor $\openone+M+{\bf C}M{\bf C}$ is a block-diagonal matrix.

We have expressed all quantities needed in terms of $| m \rangle$ and now it
remains to find $| m \rangle$ itself. For that purpose we rewrite equations
(\ref{eq:dress-eq}), (\ref{eq:dress-eqA}) in the form:
\begin{equation}\label{eq:dress-eqL1}
\begin{split}
\Phi(x,t,\lambda)\left(\ri\partial_x +\lambda
L_1^{(0)}\right)\Phi^{-1}(x,t,\lambda) &=\lambda L_1^{(1)}\\
\Phi(x,t,\lambda)\left(\ri\partial_t + \sum^N_{k=1}\lambda^k
A_k^{(0)}\right)\Phi^{-1}(x,t,\lambda)& = \sum^N_{k=1}\lambda^k A_k^{(1)}.
\end{split}\end{equation}
It is obviously satisfied at $\lambda=0$. After equating the residues of (\ref{eq:dress-eqL1}) at $\lambda= \mu^*$ to $0$ we obtain a set of
differential equations
\begin{equation}\label{eq_m}\begin{split}
\left(\openone+\frac{\mu^* M}{\mu^*- \mu} + \frac{\mu^* {\bf
C}M{\bf C}}{\mu^* + \mu}\right)\left(\ri\partial_x
+\mu^* L_1^{(0)}\right)| m \rangle=0\\
\left(\openone + \frac{\mu^* M}{\mu^*- \mu} + \frac{\mu^* {\bf
C}M{\bf C}}{\mu^* + \mu}\right)\left(\ri\partial_t + \sum^N_{k=1}
(\mu^*)^kA_k^{(0)}\right)| m \rangle=0.
\end{split}\end{equation}
Taking into account (\ref{eq:dress-m}) the equations above can be reduced to
\begin{equation}\label{pde_m}\begin{split}
\left(\ri\partial_x +\mu^* L_1^{(0)}(x,t)\right)| m (x,t)\rangle
= h(x,t)| m (x,t)\rangle\\
\left(\ri\partial_t + \sum^N_{k=1} (\mu^*)^kA_k^{(0)}(x,t)\right)| m (x,t)\rangle
=h(x,t)| m (x,t)\rangle
\end{split}\end{equation}
for some arbitrary function $h$. At this point we recall that the vectors in the decomposition (\ref{M_decomp})are not uniquely determined. Indeed, the operation
$| n \rangle \to B^{-1}| n \rangle$ and $| m \rangle \to B^{\dag}| m \rangle$ for
any nondegenerate $3\times 3$ matrix $B$ produces another decomposition
of $M$. It is not hard to see that it is always possible to choose $B$ in such a
way that $h\equiv 0$ is fulfilled. Thus from (\ref{pde_m}) it follows that
$| m (x,t)\rangle$ is proportional to some fundamental solution $\psi_0(x,t,\lambda)$
of the bare linear problem, namely
\begin{equation}
\label{m_psi}
| m (x,t)\rangle= \psi_0(x,t,\mu^*)| m_0\rangle,
\end{equation}
where $| m_0 \rangle\in\bbbc^3\backslash \{0\}$ is a constant vector of
integration. The new solution $L^{(1)}_1$ of (\ref{nlee}) and the solution
$\psi_1(x,t,\lambda)$ of the corresponding linear system are parameterized by
a complex number $\mu$ and a complex $3$-vector $| m_0\rangle$.

Thus we have proved the following Proposition:
\begin{proposition}
Let $L^{(0)}_{1}$ be a solution of (\ref{nlee}) and $\psi_0(x,t,\lambda)$
be a common solution to (\ref{bare_lax}). Let also $\mu$ be a
complex number to fulfill $\Re\mu\neq 0$, $\text{Im}\; \mu > 0$ and $| m_0\rangle\in \bbbc^3\backslash\{0\}$. Then the matrix-valued function $L^{(1)}_1(x,t)$
defined by (\ref{L_1}) where $M=|n\rangle\langle m|$ is determined by (\ref{n})
and (\ref{m_psi}) is a solution to (\ref{nlee}) as well. The
corresponding fundamental solution $\psi_1(x,t,\lambda)$ of (\ref{dressed_lax})
is given by $\psi_1=\Phi\psi_0$ where $\Phi(x,t,\lambda)$ is determined by (\ref{eq:dress-anz1}), (\ref{M_decomp}), (\ref{n}) and (\ref{m_psi}). $\Box$
\end{proposition}

Let us now consider the case when the poles of the dressing factor are imaginary,
i.e. we have:
\begin{equation}
\Phi(x,t,\lambda)=\openone + \lambda\left(\frac{M(x,t)}{\lambda - \ri\kappa}
+\frac{\mathbf{C}M(x,t)\mathbf{C}}{\lambda + \ri\kappa} \right),\qquad \kappa\neq 0.
\label{eq:dress-anz1a}\end{equation}
Then $\Phi^{-1}$ has the same poles as $\Phi$ and therefore the equality $\Phi\Phi^{-1}=\openone$ already contains second order poles. In this case
the natural requirement of vanishing of the matrix coefficients before
$(\lambda - \ri\kappa)^{-2}$ and $(\lambda - \ri\kappa)^{-1}$ leads to the
algebraic relations:
\begin{eqnarray}
M\mathbf{C}M^{\dag} = 0\label{algsys_m2}\\
\left(\openone + M + \frac{\mathbf{C}M\mathbf{C}}{2}\right) \mathbf{C}M^{\dag}\mathbf{C} + M\left(\openone + \mathbf{C}M^{\dag}\mathbf{C} + \frac{M^{\dag}}{2}\right) = 0.
\label{algsys_m1}\end{eqnarray}
As before in order to obtain a nontrivial result $M$ is required to be a degenerate
matrix, i.e. decomposition (\ref{M_decomp}) holds true. Then relation (\ref{algsys_m2})
is rewritten as
\begin{equation}
\langle m |\mathbf{C}| m \rangle = 0.
\label{algsys_m2a}\end{equation}
Relation (\ref{algsys_m1}) in its turn can be easily reduced to the following linear
system for 3-vector $| n \rangle$
\begin{equation}
\left(\openone + \frac{\mathbf{C}| n \rangle\langle m |\mathbf{C}}{2}
\right)\mathbf{C}| m\rangle = \ri\sigma | n \rangle.
\label{algsys_m1a}\end{equation}
by introducing some auxiliary real function $\sigma$. That linear system
allows one to express $| n \rangle$ through $\langle m |$ and $\sigma$, namely:
\begin{equation}
| n \rangle = \left(\ri\sigma - \frac{\langle m |m \rangle}{2}\mathbf{C}\right)^{-1}\mathbf{C}| m\rangle.
\label{n_malpha}\end{equation}
In order to find $| m \rangle$ and $\sigma$ we turn back to the equations
(\ref{eq:dress-eqL1}). Vanishing of the second order poles in
(\ref{eq:dress-eqL1}) leads to the conclusion that
\begin{equation}
| m (x,t)\rangle = \psi_0(x,t,-\ri\kappa)| m_{0}\rangle, \label{m_d}
\end{equation}
where $| m_{0}\rangle$ is a constant nonzero $3$-vector. After substituting (\ref{m_d})
into (\ref{algsys_m2a}) and taking into account (\ref{red1}) one convinces
himself that the components of the polarization vector $| m_0 \rangle$
are no longer independent but satisfy the constraint:
\begin{equation}
\langle m_0|\mathbf{C}| m_0\rangle = 0\qquad\Leftrightarrow\qquad
|m_{0,1}|^2 = |m_{0,2}|^2 + |m_{0,3}|^2.
\label{algsys_m2b}\end{equation}

The vanishing condition of the first order poles leads to some differential
constraint on $\sigma(x,t)$ which is integrated to give:
\begin{equation}
\sigma(x,t) = -\kappa\langle m_0|\psi^{-1}(x,t,\ri\kappa)\dot{\psi}_0(x,t, \ri\kappa)\mathbf{C}| m_{0}\rangle + \sigma_0,
\label{sigma_d}\end{equation}
where $\sigma_0\in\bbbr$ is a costant of integration.

Thus to calculate the soliton solution itself one just substitutes the result
for $ |n\rangle$ and $| m\rangle$ into $M$ and uses formula (\ref{L_1}). As it
is seen the new solution is parametrized by the polarization vector $| m_0\rangle$,
the real number $\sigma_0$ and the pole $\ri\kappa$. All this can be formulated
in the following manner:
\begin{proposition}
Let there be given a solution $L^{(0)}(x,t)$ to (\ref{nlee}), a common solution $\psi_0(x,t,\lambda)$ to (\ref{bare_lax}),
real numbers $\kappa > 0$, $\sigma_0$
and a complex nonzero vector $| m_0\rangle$ satisfying (\ref{algsys_m2b}). Then the
function $L^{(1)}_1(x,t)$ determined by (\ref{L_1}), (\ref{M_decomp}), (\ref{n_malpha}), (\ref{m_d}) and (\ref{sigma_d}) is a solution of the system
(\ref{nlee}) too. The solution $\psi_1(x,t,\lambda)$ of the dressed linear system (\ref{dressed_lax}) is given by $\psi_1=\Phi\psi_0$ where $\Phi$ is defined by (\ref{eq:dress-anz1a}), (\ref{M_decomp}), (\ref{n_malpha}), (\ref{m_d}) and
(\ref{sigma_d}).
\end{proposition}

One can apply the dressing procedure repeatedly to build a sequence of exact
solutions
\begin{equation}
L^{(0)}_1 \stackrel{\Phi_1}{\longrightarrow}L^{(1)}_1 \stackrel{\Phi_2}{\longrightarrow}\ldots\stackrel{\Phi_{N}}
{\longrightarrow}L^{(N)}_1  .
\label{recurs_sol}\end{equation}
More precisely this alternative procedure will be explained in Section 4.

\subsection{Soliton Solutions}

Let us apply the dressing procedure to the following seed solution
\begin{equation}
L^{(0)}_1(x,t) = \left(\begin{array}{ccc}
0 & 0 & 1 \\ 0 & 0 & 0 \\
1 & 0 & 0 \end{array}\right)
\label{vac}\end{equation}
of equation (\ref{nlee}). In this case a fundamental solution to (\ref{bare_lax})
reads:
\begin{equation}
\label{psi0}
\psi_0(x,t,\lambda) = \left(\begin{array}{ccc}
\cos (\lambda x+ \mathbf{f}_1(\lambda)t) \re^{\frac{\ri \mathbf{f}_0(\lambda) t}{3}}
& 0 & \ri\sin (\lambda x+ \mathbf{f}_1(\lambda)t) \re^{\frac{\ri \mathbf{f}_0(\lambda)t}{3}}\\
0 & \re^{\frac{-2\ri \mathbf{f}_0(\lambda) t}{3}} & 0 \\
\ri\sin (\lambda x+ \mathbf{f}_1(\lambda)t) \re^{\frac{\ri \mathbf{f}_0(\lambda) t}{3}} &
0 & \cos (\lambda x+ \mathbf{f}_1(\lambda)t)\re^{\frac{\ri \mathbf{f}_0(\lambda) t}{3}}
\end{array}\right).
\end{equation}
We recall that $\mathbf{f}_0(\lambda)$ and $\mathbf{f}_1(\lambda)$ are even and odd part of the
dispersion law induced by the $\bbbz_2$ grading of $\mathfrak{sl}(3)$, see
(\ref{f_split}).

We are going to consider the generation of a quadruplet soliton first. In this
case one uses factor (\ref{eq:dress-anz1}). It is convenient to decompose the polarization vector $|m_0\rangle $ according to the eigensubspaces of the endomorphism $\psi_0$ (\ref{psi0}):
\begin{equation}
\label{m0}| m_0\rangle=\alpha \left(\begin{array}{r}
1\\0\\1 \end{array}\right)+\beta \left(\begin{array}{r}
1\\0\\-1 \end{array}\right)+\gamma\left(\begin{array}{r}
0\\1\\0 \end{array}\right),
\end{equation}
where $\alpha,\beta,\gamma$ are arbitrary complex constants.

If the vector $| m_0\rangle$ is proportional to one of the eigenvectors of the
endomorphism $\psi_0$, then the corresponding matrix $M$ does not depend on the
variables $x$ and $t$ (due to the projective nature of the vector $| m\rangle$)
and the corresponding solution (\ref{L_1}) is a simple unitary rotation of the
constant solution $L^{(0)}_1$.

Thus elementary solitons correspond to vectors $| m_0\rangle$, belonging to
essentially two-dimensional invariant subspaces of $\psi_0$, i.e. they
correspond to polarization vectors with only one zero coefficient in the
expansion (\ref{m0}). Let us consider each of these three cases in more detail.

\subsubsection*{Case (i):  $\alpha\ne 0$, $\beta\ne 0$, $\gamma=0$}

\begin{figure}[t]
\centering
\includegraphics[width=0.4\textwidth]{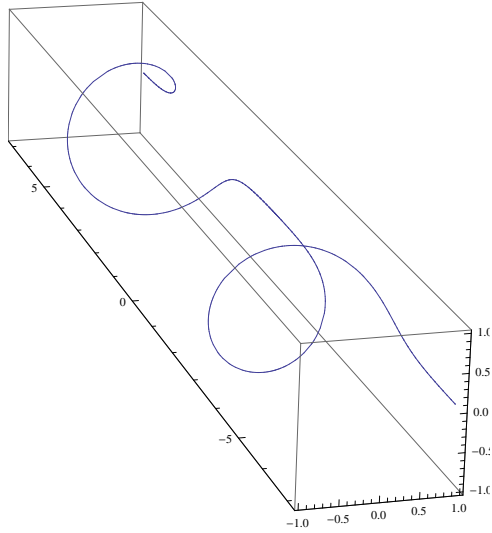}
\caption{Real and imaginary parts of the component $v_1$ in \ref{solugamma0}
as a function of $x$. Here $\kappa=1,\omega=10^{-3},\alpha =1,\beta=1+\ri$.}
\label{fig:gamma03d}
\end{figure}

The $1$-soliton solution is given by:
\begin{equation}
\begin{split}
u_1(x,t) &= 0\\
v_1(x,t) &= \exp\left\{4\ri\arctan\left(\frac{\kappa\cos(2\omega x + 2\mathbf{f}^{R}_1(\mu)t + \phi_{\alpha} - \phi_{\beta})}{\omega\cosh(2\kappa x+ 2\mathbf{f}^{I}_1(\mu)t + \ln|\alpha/\beta|)}\right)\right\},
\end{split}
\label{qsol_1}\end{equation}
where $\phi_\alpha={\rm arg}\,\alpha$, $\phi_\beta={\rm arg}\,\beta$. $\mathbf{f}^{R}_1(\lambda)$
and $\mathbf{f}^{I}_1(\lambda)$ are the real and the imaginary part of the polynomial $\mathbf{f}_1(\lambda)$
(resp. $\mathbf{f}^{R}_0(\lambda)$ and $\mathbf{f}^{I}_0(\lambda)$ stand for the real and imaginary part of
$\mathbf{f}_0(\lambda)$ to be used later on). If the dispersion law of NLEE is an even polynomial
($\mathbf{f}_1(\lambda) \equiv 0$) then the solution (\ref{qsol_1}) becomes stationary:
\begin{equation}
\label{solugamma0}
\begin{split}
u_1(x,t) &= 0\\
v_1(x,t) &= \exp\left\{4\ri\, \arctan\left(\frac{\kappa\cos(2 \omega
x+\phi_\alpha-\phi_\beta)}{\omega\cosh(2\kappa x+\ln|\alpha/\beta|)}\right)\right\}.
\end{split}
\end{equation}
A plot of that solution is presented on Fig.\ref{fig:gamma03d}. It is easy
to check that $u=0$, $v=\exp(\ri f(x))$ is an exact solution of
(\ref{nee}) for any differentiable function $f(x)$ tending to $0$ when
$x\to\pm\infty$. This resembles the case of the three-wave equation \cite{zman-3w}
where one wave of an arbitrary shape is an exact solution of the system and
the two other waves are identically zero. The solution (\ref{solugamma0}) has
a simple spectral characterisation and an explicitly given analytic fundamental
solution of the corresponding linear problem.

If the dispersion law contains odd powers of $\lambda$ as well then the
elementary soliton is no more stationary. For example in the case of equation
(\ref{nee2}) it reads:
\begin{equation}
\begin{split}
u_1(x,t) = 0,\qquad v_1(x,t) = \exp(4\ri\arctan\zeta_{\rm cub}(x,t))\\
\zeta_{\rm cub}(x,t)  = \left[\frac{\kappa\cos 2\omega [x
+ 8(3\kappa^2-\omega^2)t + (\phi_{\alpha} - \phi_{\beta})/2\omega]}
{\omega\cosh 2\kappa [x+ 8(\kappa^2-3\omega^2)t + \ln|\alpha/\beta|/2\kappa]}\right].
\end{split}
\label{qsol_1_odd}\end{equation}

\subsubsection*{Case (ii): $\alpha\ne 0$, $\beta= 0$, $\gamma\ne 0$}

\begin{figure}[t]
\centering
\includegraphics[width=0.95\textwidth]{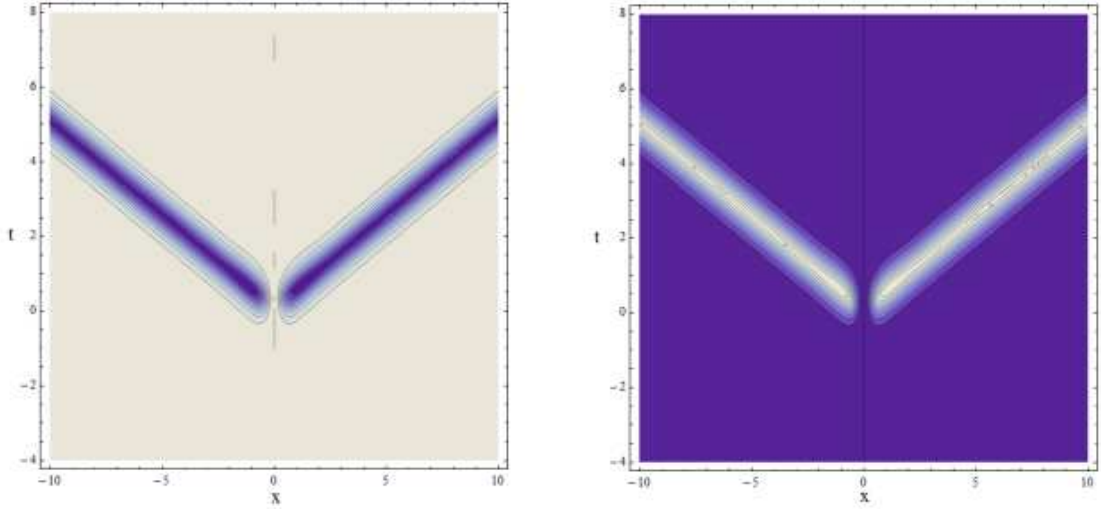}

\caption{Contour plot of $|u_1|^2$ (left panel) and $|v_1|^2$ (right panel) for a generic soliton solution (\ref{solbeta0}) as a function of
$x$ and $t$ where $\alpha=\gamma=\kappa=\omega=1$.
\label{fig:sol-1}}       
\end{figure}

In this case the solution looks as follows:
\begin{equation}\label{q_sol2}
\begin{split}
u_1(x,t) &= \frac{4\ri\omega\kappa Q^*_{{\rm gen}}\exp\ri\{\omega x+ (\mathbf{f}^{R}_0(\mu)
+ \mathbf{f}^{R}_1(\mu)) t + \phi_\alpha - \phi_\gamma\}}{(\omega-\ri\kappa)Q^2_{{\rm gen}}}\\
v_1(x,t) &= 1-\frac{8\omega\kappa^2}{(\omega-\ri\kappa)Q^2_{{\rm gen}}}\quad ,
\end{split}
\end{equation}
where $\phi_\alpha  = \arg \alpha$, $\phi_\gamma = \arg \gamma$ and
\[ Q_{{\rm gen}}=2\omega \re^{\kappa x + (\mathbf{f}^{I}_0(\mu) + \mathbf{f}^{I}_1(\mu))t
+\ln |\alpha/\gamma|} +(\omega+\ri\kappa)\re^{-\kappa x
- (\mathbf{f}^{I}_0(\mu) + \mathbf{f}^{I}_1(\mu))t -\ln |\alpha/\gamma|}.\]
In particular, when $\mathbf{f}(\lambda) = -\lambda^2 I$, i.e. $\mathbf{f}_0(\lambda) = -\lambda^2$ and $\mathbf{f}_1(\lambda) =0$ hold, we obtain a solution to (\ref{nee}):
\begin{equation}\label{solbeta0}
\begin{split}
u_1(x,t) &= \frac{4\ri\omega\kappa Q^*\exp\ri\{\omega x + (\kappa^2-\omega^2) t
+\phi_\alpha - \phi_\gamma\}}{(\omega-\ri\kappa)Q^2}\\
v_1(x,t) &= 1-\frac{8\omega\kappa^2}{(\omega-\ri\kappa)Q^2},
\end{split}
\end{equation}
where
\[ Q=2\omega \re^{\kappa(x-2\omega t)+\ln |\alpha/\gamma|}
+(\omega+\ri\kappa)\re^{-\kappa(x-2\omega t)-\ln |\alpha/\gamma|}.
\]
Contour plots of $|u_1|^2$  and $|v_1|^2$ of the solutions (\ref{solbeta0})
are shown on Figure \ref{fig:sol-1}.

When the dispersion law is odd, say ${\bf f}_1(\lambda) = -8\lambda^3$, the
quadruplet solution represents a traveling wave of the form:
\begin{equation}\label{q_sol2_odd}
\begin{split}
u_1(x,t) &= \frac{4\ri\omega\kappa Q^*\exp\ri\omega
[x+ 8(3\kappa^2-\omega^2)t + (\phi_\alpha - \phi_\gamma)/\omega]}
{(\omega-\ri\kappa)Q^2}\\
v_1(x,t) &= 1-\frac{8\omega\kappa^2}{(\omega-\ri\kappa)Q^2},\quad ,
\end{split}
\end{equation}
where
\[ Q=2\omega \re^{\kappa (x + 8(\kappa^2-3\omega^2)t
+\ln |\alpha/\gamma|/\kappa)} +(\omega+\ri\kappa)\re^{-\kappa (x
+ 8(\kappa^2-3\omega^2)t + \ln |\alpha/\gamma|/\kappa)}.\]
This is an elementary soliton for the cubic flow NLEE (\ref{nee2}).

\subsubsection*{Case (iii): $\alpha= 0,\beta\ne 0,\gamma\ne 0$}
The solution now can be obtained from the solution in the case (ii), by
changing $\alpha\to\beta$ and $x\to -x$.

In the cases (ii) the solution (\ref{solbeta0}) is a soliton of width
$1/\kappa$ moving with velocity $2\omega$. The corresponding soliton in the
case (iii) moves with a velocity $-2\omega$.

In the generic case, when all three constants are non-zero, the solution
represents a nonlinear deformation of the above described solitons.
For $\kappa>0$ it may be viewed as a decay of unstable time independent soliton from the case
(i) into two solitons, corresponding to the cases (ii) and (iii) (see fig \ref{fig:sol-1}). For $\kappa<0$,
the solution is a fusion of two colliding solitons into a stationary one.

Let us now consider dressing by a factor with two imaginary poles (doublet case),
i.e. $\mu = \ri\kappa$. There are two essentially different cases.

\subsubsection*{Case (i): $\alpha\ne 0$, $\beta\neq 0$, $\gamma = 0$}
\begin{figure}[t]
\centering
\includegraphics[width=0.95\textwidth]{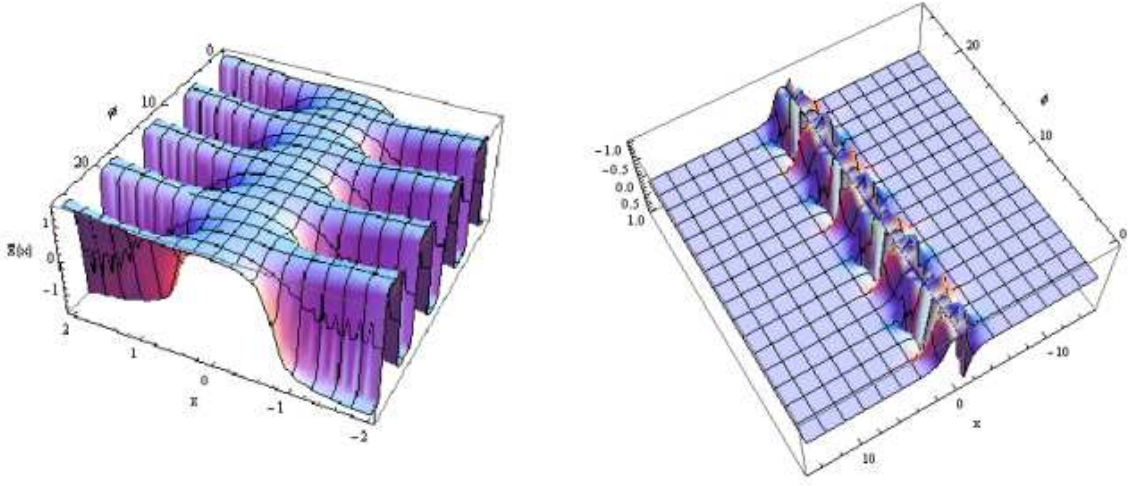}
\caption{Plots of the argument (left panel) and ${\rm Im}\, v_1(x)$  (right panel) for
the stationary solution (\ref{eq:kink} ) as a function of $x$ and $\varphi$;
$\kappa=\sigma=1$.
\label{fig:sol-3}}       
\end{figure}

From (\ref{algsys_m2b}) it follows that $|m_{0,1}| = |m_{0,3}|$. It suffices
to pick up $m_{0,1} = 1$ and the third component is $m_{0,3} =
\exp(\ri\varphi)$, $\varphi\in\bbbr$. The doublet solution reads:
\begin{equation}
\begin{split}
u_1(x,t) = 0,\qquad v_1(x,t) = \exp(4\ri\arctan\Xi_{\rm gen}(x,t))\\
\Xi_{\rm gen}(x,t) = \frac{\sigma_0 - 2\kappa (x + \dot{\mathbf{f}}_1(\ri\kappa)t)\sin\varphi}{\cosh 2(\kappa x + \mathbf{f}^{I}_1(\ri\kappa)t)
+ \sinh 2(\kappa x + \mathbf{f}^{I}_1(\ri\kappa)t) \cos\varphi}.
\end{split}
\label{d_sol1}\end{equation}
If the dispersion law of NLEE is even polynomial, i.e. $\mathbf{f}_1(\lambda) \equiv 0$,
the 1-soliton solution becomes stationary:
\begin{equation}
\begin{split}
u_1(x,t) &= 0\\
v_1(x,t) &= \exp\left\{4\ri\,\arctan\left(\frac{\sigma_0 - 2\kappa x\sin\varphi}
{\cosh 2\kappa x + \sinh 2\kappa x \cos\varphi}\right)\right\}.
\end{split}
\label{eq:kink}\end{equation}
Figure \ref{fig:sol-3} presents the argument and the imaginary part of $v_1(x)$
in the stationary case as functions of $x$ and the phase $\varphi$.

As in the quadruplet case if the dispersion law is an odd polynomial the doublet
solution is time-depending. Let us consider the simplest example
$\mathbf{f}_1(\lambda) = -8\lambda^3$ corresponding to equation (\ref{nee2}).
Now (\ref{d_sol1}) obtains the form:
\begin{equation}
\begin{split}
u_1(x,t) = 0,\qquad v_1(x,t) = \exp (4\ri\arctan\Xi_{\rm cub}(x,t))\\
\Xi_{\rm\, cub}(x,t) = \frac{\sigma_0 - 2\kappa (x + 24\kappa^2t)\sin\varphi}{\cosh 2\kappa (x + 8\kappa^2t)+ \sinh 2\kappa (x + 8\kappa^2t) \cos\varphi}.
\end{split}
\label{d_sol1_odd}\end{equation}

\subsection*{Case ii. Generic doublet}
Now let us assume $m_{0,2} \neq 0$. For simplicity we fix $m_{0,2} = 1$.
Then the norms of $m_{0,1}$ and $m_{0,3}$ are interrelated through
\[|m_{0,1}|^2 - |m_{0,3}|^2 = 1. \]
This is why it proves to be convenient to parametrize them as follows:
\begin{equation}
m_{0,1} = \cosh\theta_0 \re^{\ri(\varphi_0 + \tilde{\varphi})},
\qquad m_{0,3} = |\sinh\theta_0| \re^{\ri(\varphi_0 - \tilde{\varphi})},
\end{equation}
where $\theta_0$, $\varphi_0$ and $\tilde{\varphi}$ are arbitrary real numbers.
Then the doublet soliton solution reads:
\begin{equation}\begin{split}
u_1(x,t) &= \frac{2\Delta^*}{\Delta^2}\re^{\ri\left(\mathbf{f}_0(\ri\kappa)t + \varphi_0\right)} \left[\sinh \theta_{+}\cos\tilde{\varphi} + \ri\sinh\theta_{-}\sin\tilde{\varphi}\right]\\
v_1(x,t) &= 1 + \frac{2(2\ri\sigma -1)}{\Delta} + \frac{4\ri\sigma(\ri\sigma -1)}{\Delta^2},
\end{split}
\label{d_sol2}\end{equation}
where
\begin{eqnarray*}
\Delta (x,t)&=& \cosh^2\theta_{+}\cos^2\tilde{\varphi}
+ \cosh^2 \theta_{-}\sin^2\tilde{\varphi} - \ri\sigma\\
\sigma(x,t) &=& \sigma_0 + \kappa \dot{\mathbf{f}}^{I}_{0}(\ri\kappa) t
 + \kappa \left(x + \dot{\mathbf{f}}_1(\ri\kappa)t\right)\sinh 2 \theta_0\sin2\tilde{\varphi}\\
\theta_{\pm}(x,t) &=& \kappa x + \mathbf{f}^{I}_1(\ri\kappa)t \pm \theta_{0}.
\end{eqnarray*}

\begin{figure}[t]
\centering
\includegraphics[width=0.95\textwidth]{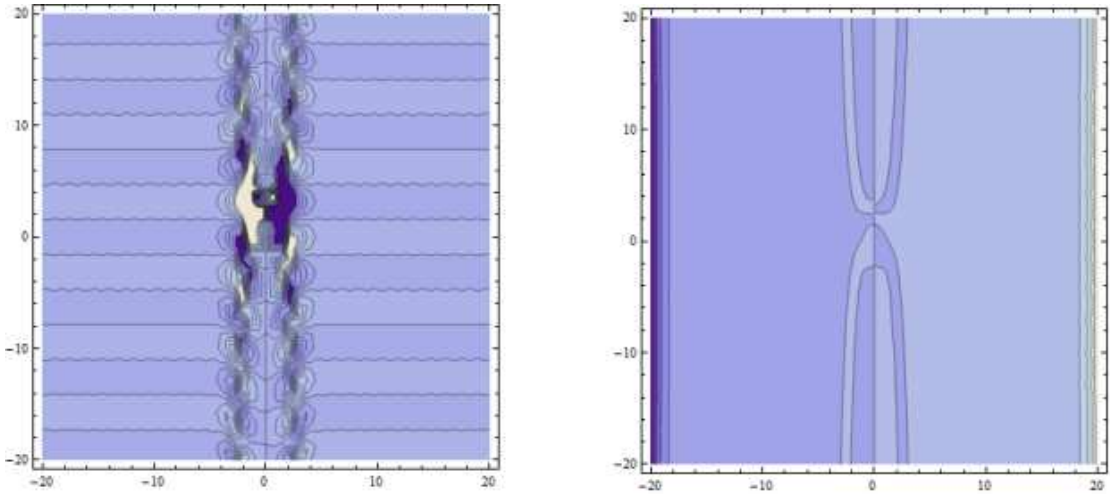}
\caption{Contour plot of $\Re\, u_1(x,t)$ (left panel) and $\Re\,v_1(x,t)$
(right panel) for doublet soliton (\ref{eq:kink2}) as functions of $x$ and $t$.
Here $\kappa=0$, $\sigma_0=5$ and $\theta_0=0$.
\label{fig:sol-4}}       
\end{figure}

Let us consider the special case when the dispersion law is $-\lambda^2 I$.
The solution (\ref{d_sol2}) is significantly simplified if in addition one
assumes that $m_{0,3}/m_{0,1}>0$ ($\tilde{\varphi} = 0$). The result reads:
\begin{equation}\begin{split}
u_1 &= \frac{2\left(\cosh^2(\kappa x + \theta_0) + \ri(\sigma_0 - 2\kappa^2 t)\right)}{\left(\cosh^2(\kappa x + \theta_0) - \ri(\sigma_0 - 2\kappa^2 t)\right)^2}\,\re^{\ri\left(\kappa^2t + \varphi_0\right)}\sinh(\kappa x + \theta_0)\\
v_1 &= \left(\frac{\cosh^2(\kappa x + \theta_0) + \ri(\sigma_0 - 2\kappa^2 t)}{\cosh^2(\kappa x + \theta_0) - \ri(\sigma_0 - 2\kappa^2 t)}\right)^2- \frac{2}{\left(\cosh^2(\kappa x + \theta_0) - \ri(\sigma_0 - 2\kappa^2 t)\right)^2}\,.\label{eq:kink2}
\end{split}\end{equation}
A plot of $\Re\, u_1(x,t)$ and $\Re\,v_1(x,t)$ is shown on Fig. \ref{fig:sol-4}

It proves to be of some interest to consider the odd dispersion case as well.
In the simplest nontrivial situation when $\mathbf{f}_1(\lambda) = -8\lambda^3$
(equation (\ref{nee2})) we have
\begin{equation}\begin{split}
u_1 &= \frac{2\left(\cosh^2(\kappa x + 8\kappa^3t + \theta_0 + \ri\sigma_0\right)}{\left(\cosh^2(\kappa x + 8\kappa^3t + \theta_0)
- \ri\sigma_0\right)^2}\,\re^{\ri\varphi_0}\sinh(\kappa x + 8\kappa^3t
+ \theta_0)\\
v_1 &= \left(\frac{\cosh^2(\kappa x + 8\kappa^3t + \theta_0)
+ \ri\sigma_0}{\cosh^2(\kappa x + 8\kappa^3t + \theta_0)
- \ri\sigma_0}\right)^2 - \frac{2}{\left(\cosh^2(\kappa x + 8\kappa^3t + \theta_0)
- \ri\sigma_0\right)^2}.\label{eq:kink2_odd}
\end{split}\end{equation}
We have assumed above that $\tilde{\varphi} = 0$.

\begin{remark}
Let us make a few short remarks on the behaviour of doublet soliton (\ref{eq:kink2}).
First of all it is evident that this {\bf is not} a travelling wave solution. Moreover, as
it is seen from Fig. \ref{fig:uv} the component $|u_1(x,t)|^2$ has two symmetric maxima
and one minimum at the origin (resp. $|v_1(x,t)|^2$ has two symmetric minima and one
maximum at the origin). The value of the maximum of $|u_1(x,t)|^2$ (resp. the minimum of $|v_1(x,t)|^2$) first increases with time ($\sigma(t) >0$) and then decreases ($\sigma(t) <0$). The maxima positions of $u_1$ depend on $t$ according to:
\begin{equation}\label{eq:}\begin{split}
\xi_0(t) =- \frac{\theta_0}{\kappa} + \frac{1}{\kappa} \ln \left( \sqrt{1+ \sqrt{1+\sigma^2 (t)}}
+\sqrt[4]{1+\sigma^2(t)}\right),
\end{split}\end{equation}
where $\sigma(t) = \sigma_0 -2\kappa^2 t$. The soliton velocity $v:=d\xi_0/dt$ {\bf is not constant} but changes with $t$ as given by:
\begin{equation}\label{eq:2}\begin{split}
v(t)= - \frac{ 2\kappa^2 t \sigma(t)}{1+\sigma^2(t)} \frac{ \sqrt[4]{1+\sigma^2(t)}}{\sqrt{1+\sqrt{1+\sigma^2(t)}}}.
\end{split}\end{equation}
Such behavior resembles the boomerons and the trappons \cite{CaDe1,CaDe2}. On Fig. \ref{fig:vel}
it is plotted the $t$-dependence of the soliton velocity.

\begin{figure}[t]
\centering
\includegraphics[width=0.45\textwidth]{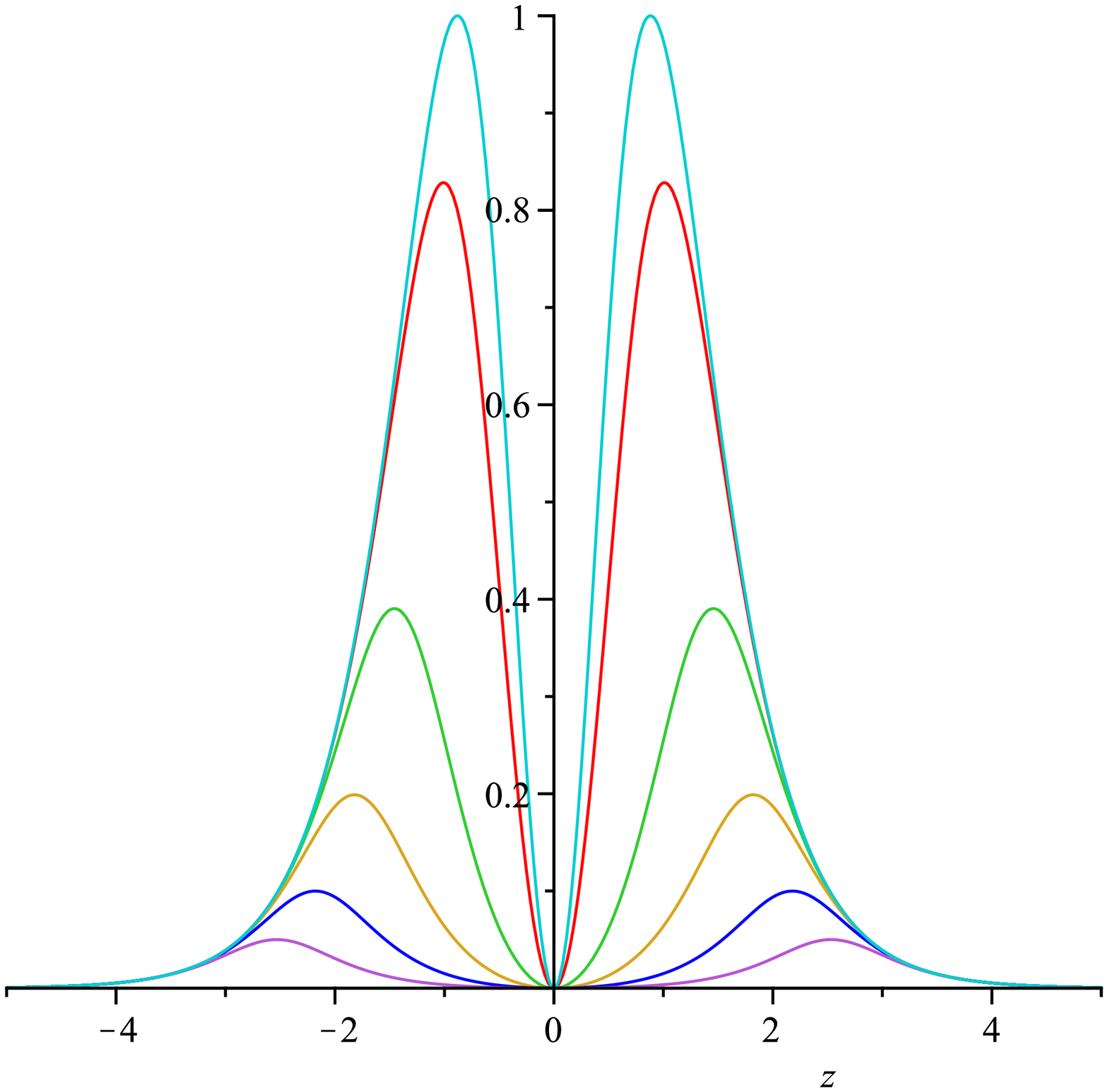}\qquad \includegraphics[width=0.45\textwidth]{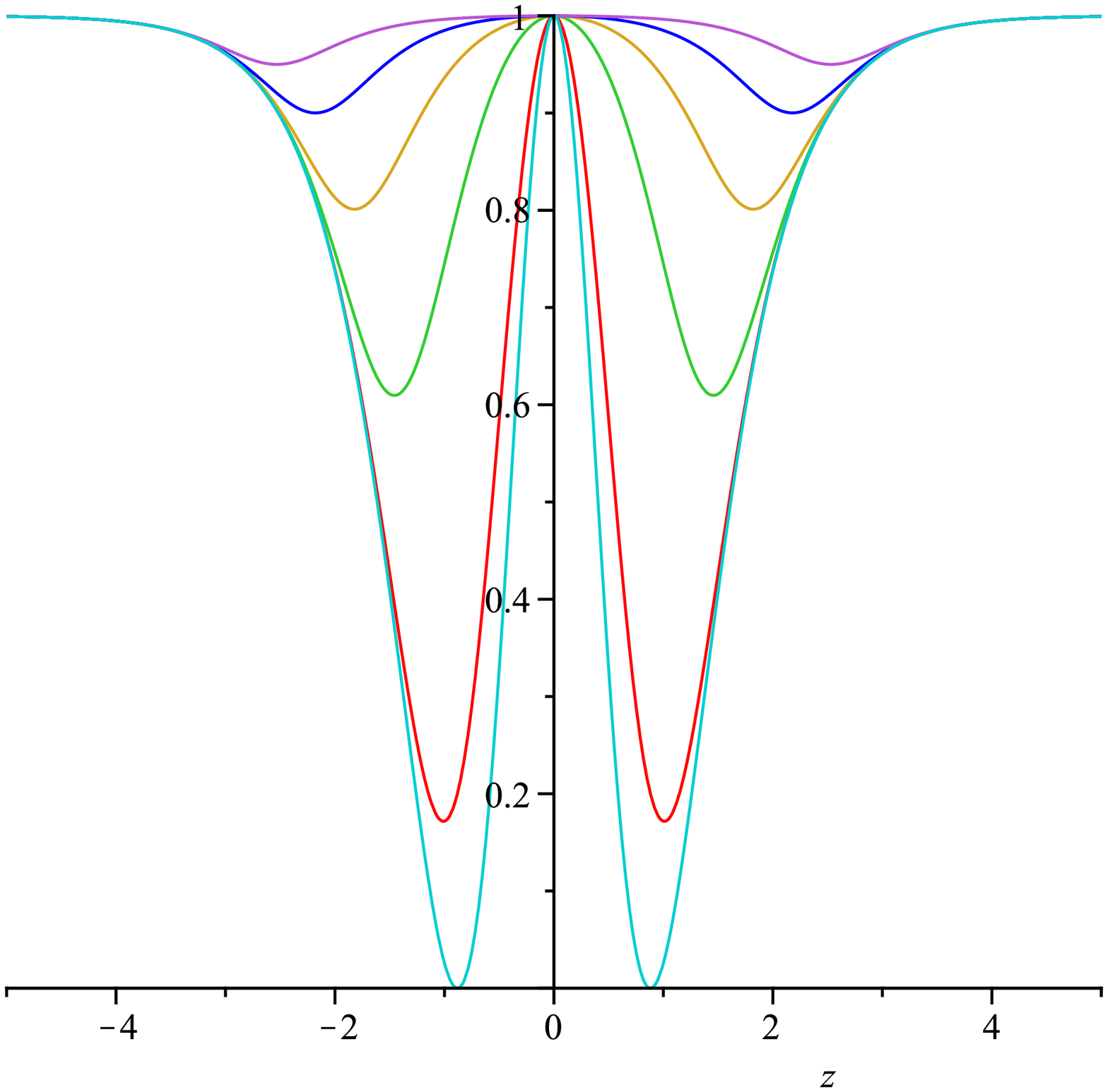}
\caption{Contour plot of $| u_1(x,t)|^2$  (left panel)  and $| v_1(x,t)|^2$ (right panel)
for doublet soliton solution  (\ref{eq:kink2}) as a function of $x$
for several values of $t$: $t=0,1,5,10,20,40$.
\label{fig:uv}}       
\end{figure}

\begin{figure}[t]
\centering
\includegraphics[width=0.45\textwidth]{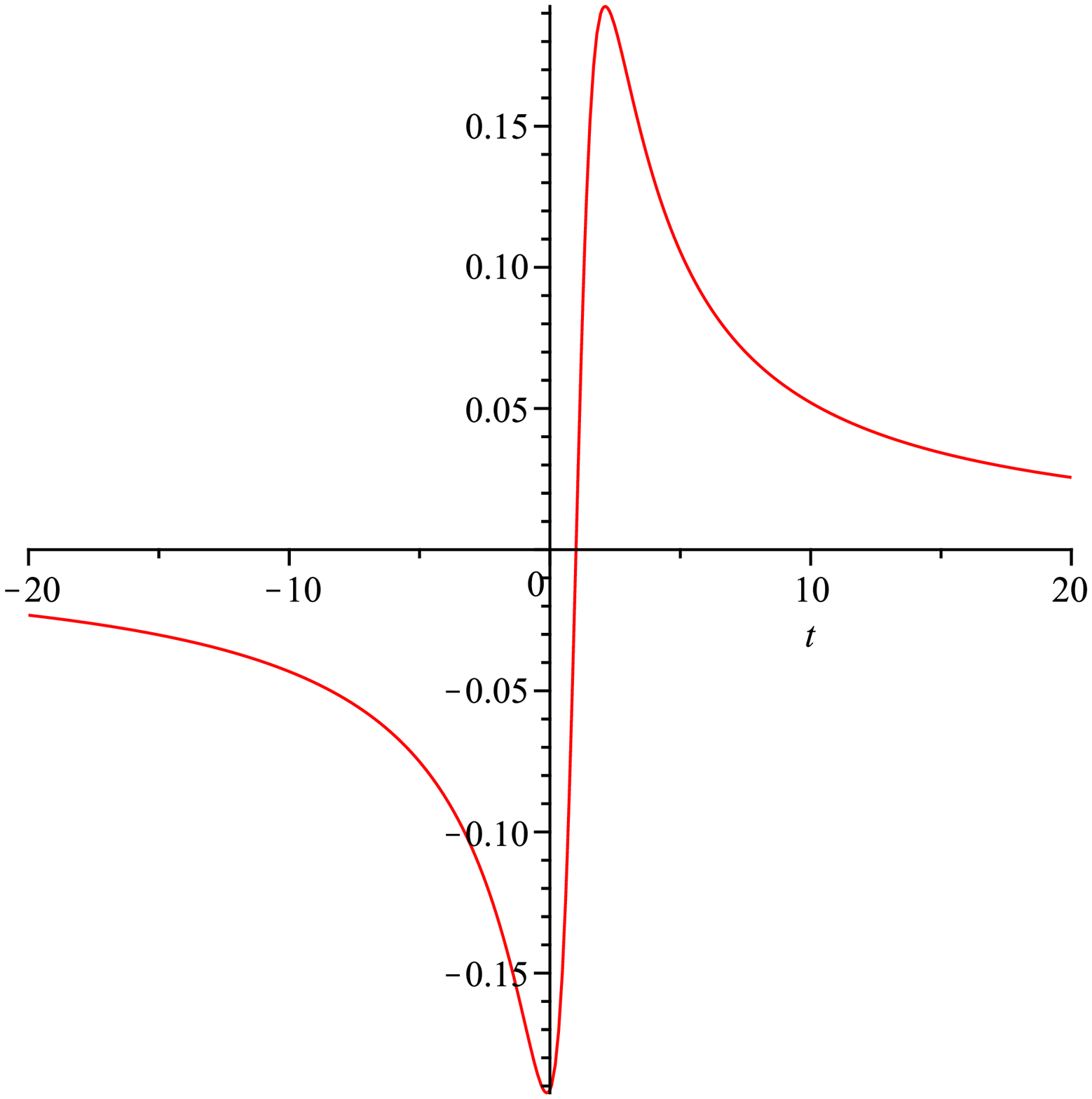}  \qquad \includegraphics[width=0.45\textwidth]{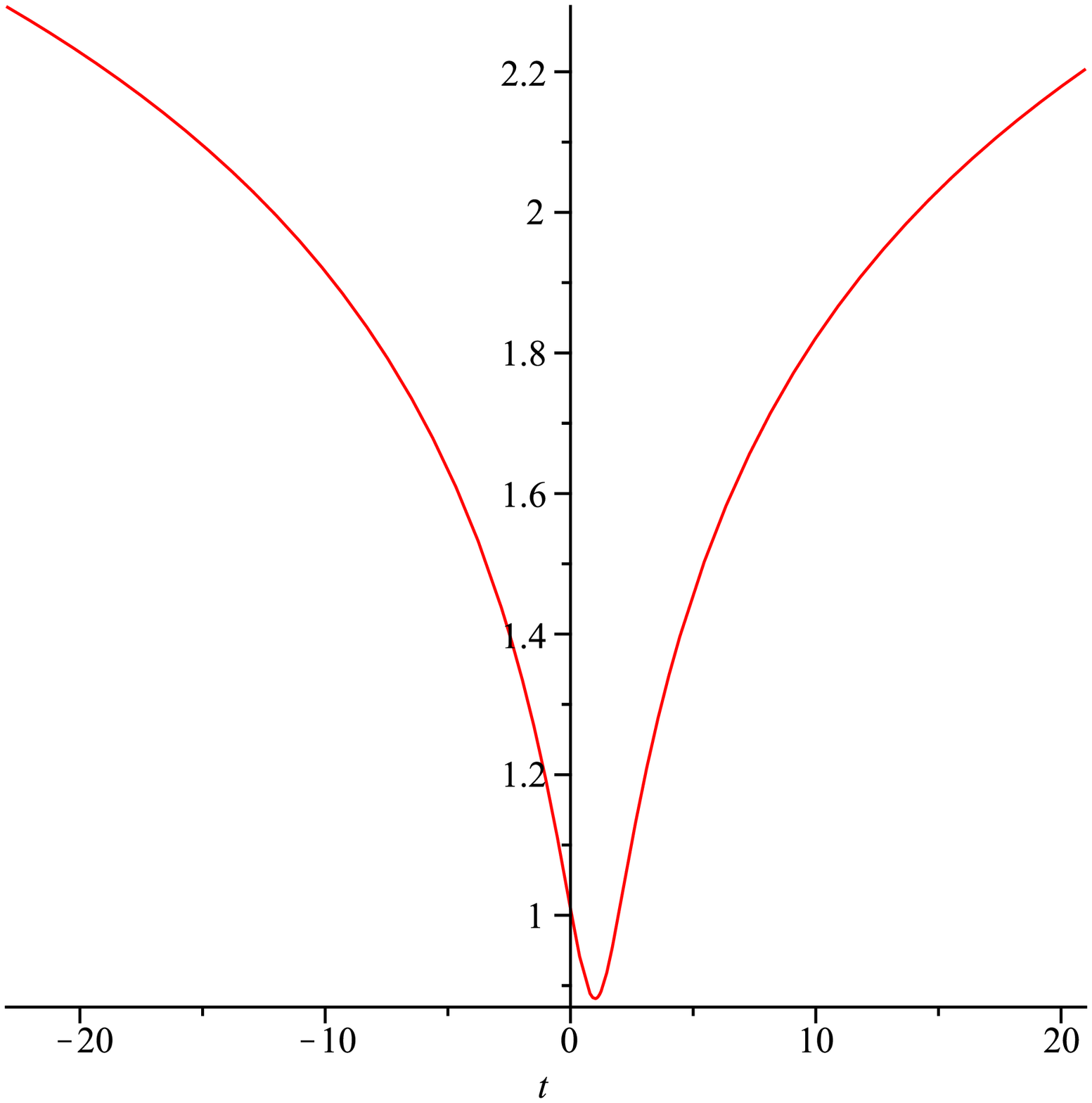}
\caption{The soliton velocity $v(t)$ and position of the maxima $\xi_0(t)$
of solution (\ref{eq:kink2}) as a functions of $t$. Here $\kappa=\sigma_0=1$,
$\theta_0=0$ and $\tilde{\varphi} = 0$.
\label{fig:vel}}       
\end{figure}

\end{remark}

\subsection{Multisoliton Solutions}

As we have already mentioned the dressing procedure can be applied several
times consequently. Thus after dressing the $1$-soliton solution one derives
a $2$-soliton solution, after dressing the $2$-soliton solution one obtains
a $3$-soliton solution and so on. Of course, in doing this one is allowed to
apply either of dressing factors (\ref{eq:dress-anz1}) and (\ref{eq:dress-anz1a}).
Therefore the multisoliton obtained will be a certain combination of quadruplet and doublet solitons.
Another way of derivation the multisoliton solution consists in using a dressing factor with a proper
number of poles:
\begin{equation}
\Phi=\openone+\lambda\sum^{N_1}_{k=1}\left(\frac{M_k}{\lambda-\mu_k}
+\frac{\mathbf{C}M_k\mathbf{C}}{\lambda+\mu_k}	\right) + \lambda\sum^{N_2}_{l=1}
\left(\frac{P_l}{\lambda-\ri\kappa_l} + \frac{\mathbf{C}P_l\mathbf{C}}{\lambda+\ri\kappa_l}\right).
\label{phi_gen}\end{equation}
As it follows from (\ref{phi_gen}) the multisoliton solution obtained will be a
mixture of $N_1$ quadruplet solitons and $N_2$ doublet ones. In order to determine
the residues of $\Phi$ one follows basically the same steps as in the  case of a
$2$-poles dressing factor. Firstly, the identity $\Phi\Phi^{-1}=\openone$ implies
that the residues of $\Phi$ and $\Phi^{-1}$ fulfill some algebraic restrictions.
For example, the condition
\begin{equation}
\lim_{\lambda\to\mu_k}(\lambda - \mu_k)\Phi\Phi^{-1} = 0,\qquad k=1,\ldots,N_1
\end{equation}
for vanishing of the residue of $\Phi\Phi^{-1}$ at $\lambda = \mu_k$ leads to the
following algebraic restrictions:
\begin{equation}
M_k\left[\openone + \mu_k\sum^{N_{1}}_{r=1}\left(\frac{M^{\dag}_r}{\mu_k-\mu_r^*}+\frac{\mathbf{C}M_r^{\dag}\mathbf{C}}
{\mu_k+\mu_r^*}\right) + \mu_k\sum^{N_2}_{l=1}\left(\frac{P^{\dag}_l}{\mu_k + \ri\kappa_l} + \frac{\mathbf{C}P_l^{\dag}\mathbf{C}}
{\mu_k - \ri\kappa_l}\right)\right]=0.
\label{algsys_n1}\end{equation}
Apart of this type of constraints we have another one originating from vanishing of the coefficients before the imaginary poles:
\begin{eqnarray}
\lim_{\lambda\to \ri\kappa_l}(\lambda - \ri\kappa_l)^2\Phi\Phi^{-1} &=& (\ri\kappa_l)^2 P_l\mathbf{C}P^{\dag}_l = 0,
\quad l=1,\ldots , N_2, \label{algsys_n2}\\
\lim_{\lambda\to \ri\kappa_l}\partial_{\lambda}[(\lambda - \ri\kappa_l)^2\Phi\Phi^{-1}]
&=& \ri\kappa_l \Theta_l\mathbf{C}P^{\dag}_l\mathbf{C} + \ri\kappa_l P_l \mathbf{C}\Theta^{\dag}_l\mathbf{C} = 0,\label{algsys_n3}
\end{eqnarray}
where
\[\begin{split}
\Theta_l &= \openone + \ri\kappa_l\sum^{N_1}_{k=1} \left(\frac{M_k}{\ri\kappa_l - \mu_k}
+ \frac{\mathbf{C}M_k\mathbf{C}}{\ri\kappa_l + \mu_k}\right) + P_l + \frac{\mathbf{C}P_l\mathbf{C}}{2}\\
& + \ri\kappa_l\sum^{N_2}_{s\neq l}\left(\frac{P_s}{\ri(\kappa_l - \kappa_s)} + \frac{\mathbf{C}P_s\mathbf{C}}
{\ri(\kappa_l + \kappa_s)} \right).
\end{split}\]
Vanishing of the rest of poles of $\Phi\Phi^{-1}$ leads to algebraic constraints
which coincide with (\ref{algsys_n1})--(\ref{algsys_n3}) due to the action of $\bbbz_2$ reductions.

Since $M_k(x,t)$ and $P_l(x,t)$ must be degenerate matrices one introduces their factorizations $M_k=| n_k\rangle\langle m_k|$ and $P_l = |q_l\rangle\langle p_l|$.
Substituting it into (\ref{algsys_n1})--(\ref{algsys_n3}) we reduce the first and the
third constraint to linear systems for $|n_k\rangle$ and $|q_l\rangle$
\begin{equation}\label{matr_sys_gen1}
\begin{split}
|m_k\rangle &= \sum^{N_1}_{r=1}\mathcal{B}_{rk}|n_{\,r}\rangle +
\sum^{N_2}_{l=1}\mathcal{D}_{sk}|q_{\,s}\rangle \\
\mathbf{C}|p_l\rangle &= \sum^{N_1}_{r=1}\mathcal{E}_{rl}|n_{\,r}\rangle +
\sum^{N_2}_{s=1}\mathcal{F}_{sl}|q_{\,s}\rangle,
\end{split}\end{equation}
where the matrix coefficients read
\begin{eqnarray*}
\mathcal{B}_{rk} &:=& \mu^*_k\left(\frac{\langle m_{\,r}|m_{k}\rangle}{\mu_r-\mu_k^*}
- \frac{\langle m_{\,r}|\mathbf{C}|m_{k}\rangle}{\mu_r+\mu_k^*}\mathbf{C}\right)\\ \mathcal{D}_{sk} &:=& \mu^*_k\left(\frac{\langle p_{\,s}|m_{k}\rangle}{\ri\kappa_s-\mu_k^*} - \frac{\langle p_{\,s}|\mathbf{C}|m_{k}\rangle}{\ri\kappa_s+\mu_k^*}\mathbf{C}\right)\\
\mathcal{E}_{rl} &:=&	-\ri\kappa_l\left(\frac{\langle m_{\,r}|\mathbf{C}|p_{l}\rangle}{\ri\kappa_l - \mu_k}
+ \frac{\langle m_{\,r}|p_{l}\rangle}{\ri\kappa_l + \mu_k}\mathbf{C}\right),
\quad \mathcal{F}_{ss} :=\ri\sigma_{s} - \frac{\langle p_{\,s}|p_{s}\rangle}{2}\mathbf{C}\\
\mathcal{F}_{sl} &:=& \kappa_l\left(\frac{\langle p_{\,s}|\mathbf{C}|p_{l}\rangle}{\kappa_s-\kappa_l}
- \frac{\langle p_{\,s}|p_{l}\rangle}{\kappa_s+\kappa_l}\mathbf{C}\right),\qquad s\neq l
\end{eqnarray*}
By inverting the linear system (\ref{matr_sys_gen1})  we
can express $| n_{\,r}\rangle$ and $| q_{\,s}\rangle$ through all $| m_{k}\rangle$,
$| p_{\,l}\rangle$ and $\sigma_l$ and that way determine the dressing factor
in terms of the latter. The vectors $| m_{k}\rangle$ and $| p_{\,l}\rangle$ as well as the functions $\sigma_l$ can be found from the natural requirement of vanishing of the poles in (\ref{eq:dress-eqL1}). The result reads
\begin{equation}\label{FGl_res}\begin{split}
|m_{k}(x,t)\rangle &= \psi_0(x,t,\mu^*_k)
| m_{k,0}\rangle\\
|p_{\,l}(x,t)\rangle &= \psi_0(x,t,-\ri\kappa_l)
| p_{\,l,0}\rangle\\
\sigma_l(x,t) &= -\kappa_l\langle p_{l,0}|\psi^{-1}(x,t,\ri\kappa_l)\dot{\psi}_0(x,t, \ri\kappa_l)\mathbf{C}| p_{l,0}\rangle + \sigma_{l,0}.
\end{split}\end{equation}

Analogously to the $2$-poles case the components of $|p_l\rangle$ are
not independent. As a result of (\ref{algsys_n2}) that the following relations
holds true:
\begin{equation}
\langle p_l(x,t) |\mathbf{C}| p_l(x,t)\rangle
= \langle p_{l,0} |\mathbf{C}| p_{l,0}\rangle = 0.
\end{equation}

Thus we have proved that the dressing factor in the multiple poles case is
determined if one knows the initial fundamental solution $\psi_{0}(x,t,\lambda)$.
The multisoliton solution itself can be derived through the following formula
\begin{equation}
L_1^{(1)}(x,t) = \Phi(x,t,\infty) L_{1}^{(0)}(x,t)\Phi^{\dag}(x,t,\infty),
\end{equation}
where
\[\Phi(x,t,\infty) = \openone + \sum^{N_1}_{k=1} (M_k + \mathbf{C}M_k\mathbf{C})
+ \sum^{N_2}_{l=1} (P_l + \mathbf{C}P_l\mathbf{C}).\]
From all said above it follows that the algorithm for obtaining the multisoliton solution
can be presented symbolically as follows
\[\begin{split}
L_{1}^{(0)}\longrightarrow \{| m_k\rangle\}^{N_1}_{k=1},\ \{| p_l\rangle\}^{N_2}_{l=1},\ \{\sigma_{l}\}^{N_2}_{l=1} \longrightarrow\\ \{| n_k\rangle\}^{N_1}_{k=1}, \{| q_l\rangle\}^{N_2}_{l=1} \longrightarrow \{M_k\}^{N_1}_{k=1},\{P_l\}^{N_2}_{l=1}
\longrightarrow L_1^{(1)}.
\end{split}\]

\section{Interactions of Quadruplet Solitons }

In this section we aim to study the interactions of solitons we have derived. We shall
restrict ourselves with quadruplet solitons for NLEEs with odd dispersion laws. This is
the simplest case since the solitons are travelling wave-type solutions. The interactions
of the other types of solitons require a special treatment and will be done elsewhere.

Our study will be based on the Zakharov-Shabat scheme \cite{brown-bible} applied to
the recursive procedure (\ref{recurs_sol}). Their
approach consists in calculating the asymptotics of generic $N$-soliton solution
for $t \to \pm\infty$ and establishing the pure elastic character of the  interactions
of generic soliton, i.e. solitons travelling at different velocities. The pure elastic character of the soliton interactions is demonstrated by the fact that for $t\to\pm\infty$ the $N$-soliton solution splits into a sum of $N$ one soliton solutions preserving its amplitudes and velocities. The only effect of the interaction consists in shifting the center of mass and the initial phase of the solitons.

The 1-soliton dressing factor corresponding to the quadruplet case  with poles at $\pm \mu_k$ is given by:
\begin{equation}\label{eq:Fik}\begin{split}
\Phi_k(x,t,\lambda ,\mu_k) &= \openone + \frac{ \lambda}{\lambda -\mu_k} M_k(x,t) + \frac{ \lambda}{\lambda +\mu_k} \mathbf{C} M_k(x,t)\mathbf{C}.
\end{split}\end{equation}
The residues $M_k(x,t) = |n_k \rangle \langle m_k |$ are
determined by the following equalities
\begin{equation}\label{eq:chiN}\begin{split}
|n_k\rangle &=  \frac{1}{\mu^*_k}\left(\frac{\langle m_k|m_k\rangle}{2\ri\kappa_k}
- \frac{\langle m_k|\mathbf{C}|m_k\rangle}{2\omega_k}\mathbf{C}\right)^{-1}
| m_k \rangle\\
|m_k (x,t)\rangle &= \psi_0(x,t,\mu_k) |m_{k 0}\rangle, \qquad |m_{k 0}\rangle = \left(\begin{array}{c}
\alpha_k + \beta_k \\ \gamma_k \\ \alpha_k -\beta_k \end{array}\right).
\end{split}\end{equation}

Let us now outline the alternative procedure for constructing the $N$-soliton solutions of the NLEE (\ref{nlee}).
The idea is to apply subsequently $N$ times the the one-soliton dressing. For simplicity we assume that all
$N$ solitons are of quadruplet type. As a result the sequence of mappings (\ref{recurs_sol}) allows us to
constructs a sequence of Lax operators with potentials $L_1^{(k)}$,  $k=1,\dots,N$ and eigenfunctions:
\begin{equation}\label{eq:chi(N)}\begin{split}
\chi^\pm_{(k)} (x,t,\lambda) &= \bPhi_{k}(x,t,\lambda,\mu_k) \bPhi_{k-1}(x,t,\lambda,\mu_{k-1}) \dots
\bPhi_{1}(x,t,\lambda,\mu_1) \\
&\times \psi_0(x,t,\lambda) \bPhi_{1,-}^\dag(\lambda,\mu_1)\dots \bPhi_{k-1,-}^\dag (\lambda,\mu_{k-1}) \bPhi_{k,-}^\dag(\lambda,\mu_k),
\end{split}\end{equation}
where
\begin{equation}\label{eq:phim}\begin{split}
\bPhi_{k,-}(\lambda,\mu_k) =\lim_{x\to -\infty} \bPhi_{k}(x,t,\lambda,\mu_k)
\end{split}\end{equation}
The dressing factors $\bPhi_{k}(x,t,\lambda,\mu_k)$ are constructed in analogy with (\ref{eq:Fik}) as follows:
\begin{equation}\label{eq:bPhik}\begin{split}
\bPhi_k(x,t,\lambda ,\mu_k) &= \openone + \frac{ \lambda}{\lambda -\mu_k} \bM_k(x,t) + \frac{ \lambda}{\lambda +\mu_k} \mathbf{C} \bM_k(x,t)\mathbf{C} \\
\bM_k(x,t) &=  \frac{1}{\mu^*_k}\left(\frac{\langle \bm_k| \bm_k\rangle}{2\ri\kappa_k}
- \frac{\langle \bm_k|\mathbf{C}| \bm_k\rangle}{2\omega_k}\mathbf{C}\right)^{-1}
| \bm_k \rangle \langle \bm_k| \\
|\bm_k\rangle &= \Phi_{k-1}(x,t,\mu_k,\mu_{k=1})  \dots \Phi_{1}(x,t,\mu_k ,\mu_1) |m_k\rangle .
\end{split}\end{equation}

Thus for the $N$-soliton potential we obtain:
\begin{equation}\label{eq:L1ns}\begin{split}
L_1^{(N)} (x,t) &= \lim_{\lambda\to\infty} \chi^\pm_{(N)} (x,t,\lambda) L_1^{(0)}. \hat{\chi}^\pm_{(N)} (x,t,\lambda)
\end{split}\end{equation}

Next we recall that we are considering NLEE with odd dispersion laws (\ref{nlee}b). Their one-soliton solutions
are traveling waves and depend on $Z_k=x-V_kt$, where $V_k =1/\kappa_k \im f_1(\mu_k) $. In particular, for the eq.
(\ref{nee2}) $\mathbf{f}_1(\lambda) =-8\lambda^3$ and $V_k =8(3\mu_k^2 -\kappa_k^2)$.
Now  let us to pick up the trajectory of the $N$-th soliton:
$Z_N\equiv x - 2 \omega_N t/3 = {\rm fixed}$ and evaluate the asymptotics of $L_1^{(N)} (x,t)$ for
$t\to\pm\infty$ for fixed $Z_N$. This will allow us to see what are the effects of the soliton interactions
on the $N$-th soliton.

In what follows we will assume that all solitons move with different velocities, i.e. $V_j \neq V_k$ for $k\neq j$.
It is natural to split the solitons in two groups:
\begin{equation}\label{eq:Solpm}\begin{split}
\mathcal{M}_+ \equiv \{ V_k \colon V_k >V_N \}, \qquad
\mathcal{M}_- \equiv \{ V_k \colon V_j <\omega_N \},
\end{split}\end{equation}
i.e., the solitons belonging to $\mathcal{M}_+$ are moving faster than the $N$-th soliton, while the ones belonging to $\mathcal{M}_-$ are slower.

Now we are able to calculate the limits of $\Phi_k(x,t,\lambda)$ for $t\to\pm\infty$ for fixed $Z_N$. To do this we firstly need to obtain the limits of the one-soliton dressing factor for $x\to\pm\infty$. It can be  verified that:
\begin{equation}\label{eq:Fipm}\begin{aligned}
 \Phi_{k,-}(\lambda,\mu_k)  := \lim_{x\to -\infty} \Phi_k(x,t,\lambda) &=
\left(\begin{array}{ccc} c_k(\lambda) &  0 & -c_k'(\lambda),\\0 & 1 & 0 \\ -c_k'(\lambda) &  0 & c_k(\lambda) \end{array}\right),\\
 \Phi_{k,+}(\lambda,\mu_k)  := \lim_{x\to\infty} \Phi_k(x,t,\lambda) &= \left(\begin{array}{ccc} c_k(\lambda) &  0 & c_k'(\lambda) \\
0 & 1 & 0 \\ c_k'(\lambda) &  0 & c_k(\lambda) \end{array}\right),
\end{aligned}\end{equation}
where
\[\quad c_k(\lambda) = \frac{ \mu_k}{\mu_k^*} \frac{ \lambda^2 -|\mu_k|^2}{\lambda^2 -\mu_k^2},
\qquad c_k'(\lambda) = -\frac{ \mu_k}{\mu_k^*} \frac{ \lambda (\mu_k -\mu_k^*)}{\lambda^2 -\mu_k^2}.\]
Note that the asymptotics $\Phi_{k,\pm} (\lambda,\mu_k) $ do not depend upon the polarization vectors
$| m_{k0}\rangle $ and that they commute for different values of $\lambda$. This allows us
to describe explicitly the $N$-soliton interactions of quadruplet solitons.

The action of $\Phi_{k,\pm}(\lambda,\mu_k)$ on the polarization vectors produces
the equalities:
\begin{equation}\label{eq:FiVec}\begin{aligned}
\Phi_{k,\pm}(\lambda,\mu_k) \left(\begin{array}{c} \alpha_k +\beta_k \\ \gamma_k \\ \alpha_k -\beta_k \end{array}\right) & =
\left(\begin{array}{c} \alpha_k^\pm +\beta_k^\pm \\ \gamma_k \\ \alpha_k^\pm -\beta_k^\pm \end{array}\right) \\
\frac{ \alpha_k^\pm}{\alpha_k} = \frac{\mu_k}{\mu_k^*} \frac{\lambda \pm \mu_k^* }{\lambda \pm \mu_k} ,
&\qquad \frac{ \beta_k^\pm}{\beta_k} = \frac{\mu_k}{\mu_k^*} \frac{\lambda \mp \mu_k^* }{\lambda \mp \mu_k} .
\end{aligned}\end{equation}

Next we have to evaluate the asymptotics of $|\bm_k (x,t) \rangle $ when $t\to \pm\infty$ along the trajectory $Z_N(x,t)={\rm const}$. This is done recursively using (\ref{eq:Fipm}). Skipping all technical details here we get:
\begin{equation}\label{eq:P1}\begin{split}
|\bm_N (x,t) \rangle &\mathop{\simeq}\limits_{t\to\infty}  \prod_{j\in \mathcal{M}_+ }^{} \Phi_+(\mu_N,\mu_j)
\prod_{j\in \mathcal{M}_- }^{} \Phi_-(\mu_N,\mu_j)|m_N(x,t)\rangle \\
|\bm_N (x,t) \rangle &\mathop{\simeq}\limits_{t\to -\infty}  \prod_{j\in \mathcal{M}_+ }^{} \Phi_-(\mu_N,\mu_j)
\prod_{j\in \mathcal{M}_- }^{} \Phi_+(\mu_N,\mu_j)|m_N(x,t)\rangle.
\end{split}\end{equation}
Then from (\ref{eq:FiVec}) and (\ref{eq:P1}) one deduces that:
\begin{equation}\label{eq:P2}\begin{aligned}
\frac{\alpha_N^+ }{\alpha} &= \prod_{k=1}^{N} \frac{\mu_k}{\mu_k^*} \prod_{j\in \mathcal{M}_+ }^{} A_{N,j}
\prod_{j\in \mathcal{M}_- }^{} B_{N,j} &\qquad
\frac{\alpha_N^- }{\alpha} &= \prod_{k=1}^{N} \frac{\mu_k}{\mu_k^*} \prod_{j\in \mathcal{M}_+ }^{} B_{N,j}
\prod_{j\in \mathcal{M}_- }^{} A_{N,j} \\
\frac{ \beta_N^+}{\beta} &=  \prod_{k=1}^{N} \frac{\mu_k}{\mu_k^*}  \prod_{j\in \mathcal{M}_+ }^{} B_{N,j}
\prod_{j\in \mathcal{M}_- }^{} A_{N,j}  &\qquad
\frac{ \beta_N^-}{\beta} &=  \prod_{k=1}^{N} \frac{\mu_k}{\mu_k^*}  \prod_{j\in \mathcal{M}_+ }^{} A_{N,j}
\prod_{j\in \mathcal{M}_- }^{} B_{N,j} \\
A_{N,j}&=\frac{\mu_N +\mu_j^* }{\mu_N +\mu_j},
&\qquad B_{N,j} &=\frac{\mu_N -\mu_j^* }{\mu_N -\mu_j}.
\end{aligned}\end{equation}
As a result we obtain that: i) the soliton interactions are purely elastic, and ii)  their effect
is shifts of the relative center of mass and the phase $\delta_N =\arg \alpha -\arg \beta$ of the solitons:
\begin{equation}\label{eq:P3}\begin{aligned}
 Z_N^{\pm} &= Z_N \mp \sum_{j\in \mathcal{M}_+ } z_{N,j} \;\pm \sum_{j\in \mathcal{M}_- } z_{N,j}\\
\delta_N^{\pm} &= \delta_N \pm \sum_{j\in \mathcal{M}_+ }^{} \phi_{N,j} \;\mp \sum_{j\in \mathcal{M}_- }^{} \phi_{N,j}\\
z_{N,j} &= \frac{1}{2\kappa_N }( \ln |  A_{N,j} | -\ln | B_{N,j} | ),\quad
\phi_{N,j} = \arg ( A_{N,j} ) -\arg ( B_{N,j} ).
\end{aligned}\end{equation}

\section{Integrals of Motion}

Here we will sketch briefly the direct method for finding integrals of motion, introduced by Drinfel'd and
Sokolov \cite{DrSok*85}. We will apply it to the system
(\ref{nee}). In order to do that it proves to be technically more
convenient to deal with the Lax pair (\ref{lax_1_g}), (\ref{lax_2_g}).
We will use the transformation $\mathcal{P}(x,t,\lambda)$ that diagonalizes simultaneously the Lax pair $\tilde{L}$ and $\tilde{A}$:
\begin{equation}\label{Lambda_mu}\begin{split}
\mathcal{L}  &= \mathcal{P}^{-1}\tilde{L}\mathcal{P}
= \ri \partial_x + \lambda J +\mathcal{L}_0
+ \frac{\mathcal{L}_1}{\lambda} + \cdots\\
\mathcal{A} &= \mathcal{P}^{-1}\tilde{A}\mathcal{P}
= \ri \partial_t + \lambda^2 I + \lambda\mathcal{A}_{-1}
+ \mathcal{A}_0 + \frac{\mathcal{A}_1}{\lambda} + \cdots.
\end{split}\end{equation}
Here all matrix coefficients $\mathcal{L}_k$, $\mathcal{A}_{-1}$ and
$\mathcal{A}_k$, $k=0,1,\hdots$ are diagonal. Using the asymptotic expansion for $\mathcal{P}(x,t,\lambda)$:
\begin{equation}
\mathcal{P}(x,t,\lambda)= \openone +\frac{p_1(x,t)}{\lambda} +
\frac{p_2(x,t)}{\lambda^2} + \cdots \label{P}\end{equation}
one can get a set of recurrence relations:
\begin{eqnarray}
U_0 + J p_1  &=& \mathcal{L}_0 + p_1 J\label{lambda_0}\\
\ri p_{1,x} + U_0 p_1 &+& Jp_2 = \mathcal{L}_1 + p_1\mathcal{L}_0 + p_2 J
\label{lambda_m1}\\
& \vdots \nonumber\\
\ri p_{k,x} + U_0 p_k &+& J p_{k+1} = \mathcal{L}_k + p_{k+1}J +
\sum^{k-1}_{m=0}p_{k-m}\mathcal{L}_m \label{lambda_mk}\\
& \vdots\nonumber
\end{eqnarray}
Here
we assume that all coefficients $p_l$ ($l=1,2,\ldots$) are
off-diagonal matrices.

\begin{figure}[t]
\centering
\includegraphics[width=0.48\textwidth]{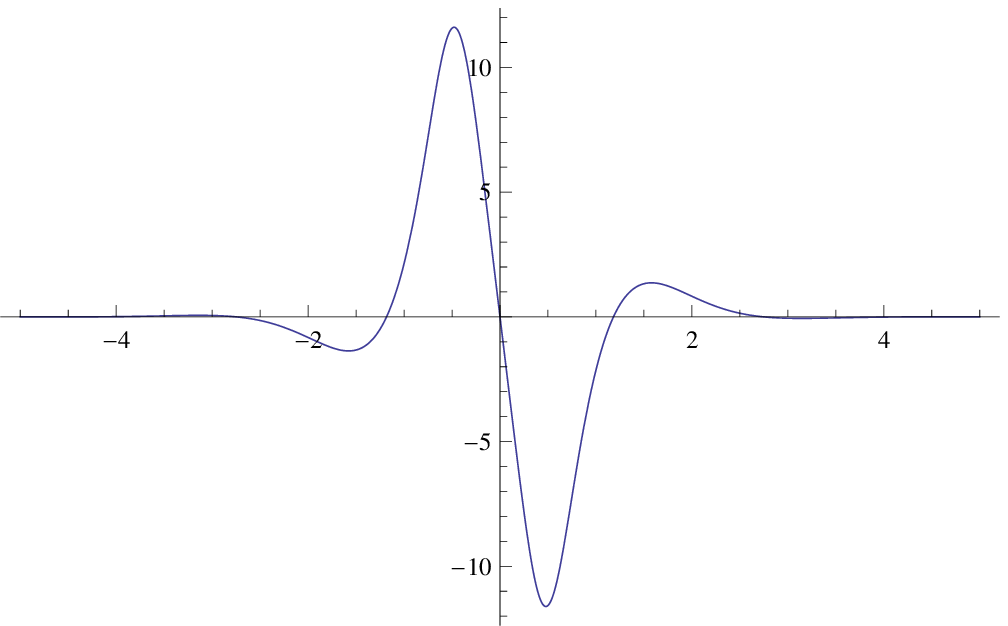} \includegraphics[width=0.48\textwidth]{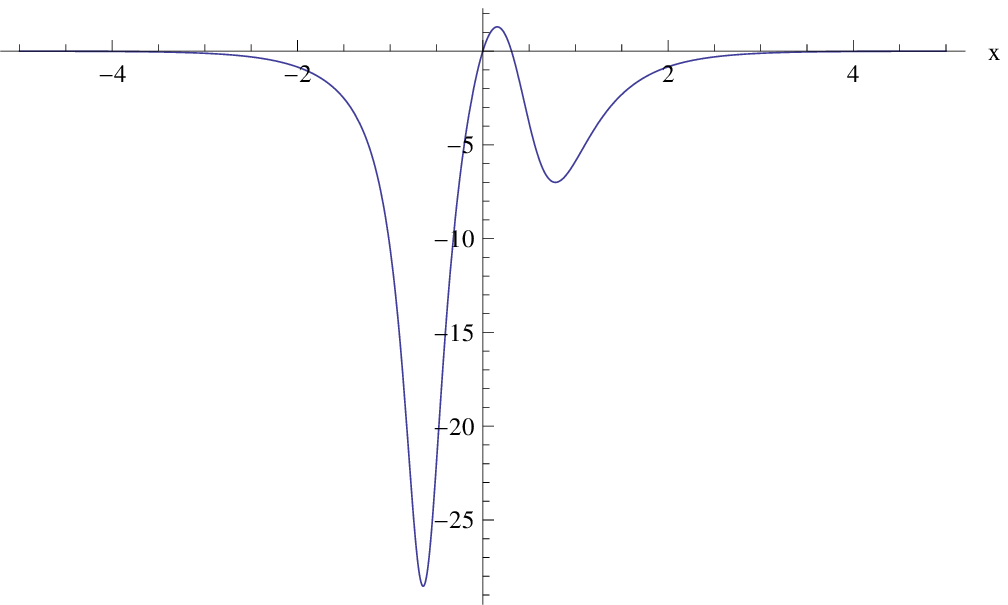}
\caption{Plots of the density of the first integral of motion as a
function of $x$  evaluated on the stationary quadruplet soliton (\ref{solugamma0}) for $\alpha =\beta=\gamma=\omega=1$, $\delta=0$
(left panel) and for the stationary doublet soliton (\ref{eq:kink}) for $\kappa=\sigma_0=1$, $\phi=0$ (right panel).}
\label{fig:integr}       
\end{figure}

In order to solve the recursion relations above, we will split
each relation into a diagonal and off-diagonal part. For example,
treating this way the first relation above one gets
\begin{equation}
\mathcal{L}_0 = U^{\rm d}_0,\qquad U^{\rm f}_0 = -[J,p_1],
\label{recursplit_1}\end{equation}
where the superscripts $d$ and $f$ above denote projection onto diagonal and
off-diagonal part of a matrix respectively. Taking into account the explicit
form of $U_0$  for $\mathcal{L}_0$ we have
\begin{equation}
\mathcal{L}_0 = \frac{\ri}{2}(uu^*_x +vv^*_x)\left(\begin{array}{ccc}
1 & 0  & 0 \\
0 & -2 & 0 \\
0 & 0  & 1
\end{array}\right).\label{Lambda_0}
\end{equation}
Thus as a density of our first integral we can choose:
$\mathcal{I}_0 = u^*u_x + v^*v_x$. It represents momentum
density of our system. For the stationary solutions (\ref{solugamma0}) and (\ref{eq:kink} the momentum density is depicted on Figure \ref{fig:integr}. It is evidential, that the integrals of motion are well localized function of $x$.

Similarly, for the second integral density one gets
\[\mathcal{I}_1=|uu^*_x+vv^*_x|^2+4|uv_x-vu_x|^2.\]

In general,  the-$k$ integral of motion can be calculated
through the formula
\begin{equation}
\mathcal{L}_k = \left(U^f_0p_k\right)^d.
\label{Lambda_k}\end{equation}
The matrix $p_k$ in its turn is obtained from the
following recursive formula
\begin{equation}
p_k = -\ad^{-1}_J\left(\ri p_{k-1,x} + (U_0 p_{k-1})^f -
\sum^{k-1}_{m=0}p_{k-1-m}\mathcal{L}_m\right).
\end{equation}

Note, that the zero curvature representation is gauge invariant, i. e.
$[\mathcal{L}, \mathcal{A}] = 0$ is fulfilled. Since
$[\mathcal{L}_k, \mathcal{A}_l] =0$ the commutativity of
$\mathcal{L}$ and $\mathcal{A}$ is equivalent to the following
requirements
\begin{equation}
\partial_x \mathcal{A}_{-1} = 0,\qquad
\partial_t\mathcal{L}_k - \partial_x \mathcal{A}_k = 0,\qquad
k=0,1,\hdots
\end{equation}
Hence $\mathcal{L}_k$ represent densities of the integrals of motion
we are interested in.

\section{Conclusions}

The soliton solutions for a hierarchy of NLEEs related to the symmetric space $SU(3)/S(U(1)$
$\times U(2))$ are constructed.  In order to obtain
the soliton solutions we have applied the dressing procedure with a 2-poles dressing
factor. It has been shown that there exist two types of 1-soliton solutions: quadruplet solitons which are associated with 4 symmetrically located eigenvalues of $L$ and doublet solitons which are associated with a pair of purely imaginary eigenvalues. This remarkable fact is
a consequence of the simultaneous action of two $\bbbz_2$ reductions on the Lax pair.
  The properties of the elementary solitons depend crucially upon
the symmetry properties the dispersion law. For example, if the dispersion law is an
even polynomial then the elementary soliton of the first type will be stationary (see formula (\ref{solugamma0})) otherwise it is time-dependent (formula (\ref{qsol_1_odd})). In the case of the doublet type solitons the situation changes significantly --- the components of the polarization vector $|m_0\rangle$ are no longer independent, see
(\ref{algsys_m2b}). This is why we have only two cases possible: generic case and a
degenerate case. In the latter case the doublet soliton is stationary if
$\mathbf{f}(\lambda)$ is an even polynomial, otherwise they are time-depending.
In the generic case a new phenomenon arises. When the dispersion law is an even
polynomial the soliton is not a traveling wave. Its behavior resembles that of
trappons and boomerons --- the soliton velocity is not fixed but varies with time.

We have described the quadruplet soliton interactions for NLEE with odd dispersion laws by calculating explicitly their
asymptotics along the soliton trajectories in the generic case (different soliton
velocities). The important result consisted in the following:

i) the $N$-soliton interactions are purely elastic and always split into sequences of elementary 2-soliton interactions;

ii) the effect of each 2-soliton interaction consists in shifts of the relative center of mass and relative
phases of each of the solitons;

iii) the corresponding shifts are different from the ones for the NLS and Heisenberg ferromagnetic equations.

\section*{Acknowledgements}

 The authors acknowledge support from the Royal Society and
the Bulgarian academy of sciences via joint research project
"Reductions of Nonlinear Evolution Equations and Analytic
Spectral theory".  The work of G.G.G. is supported by the
Science Foundation of Ireland (SFI), under grant No. 09/RFP/MTH2144.

\label{last}
\end{document}